\documentclass[preprint,12pt]{elsarticle}
\usepackage{amssymb}
\usepackage{amsmath}
\usepackage{graphicx,color}
\usepackage{algorithmic}
\usepackage{algorithm}
\usepackage{physics}
\usepackage{geometry}
\usepackage{graphicx}
\usepackage{subcaption}
\usepackage[hidelinks]{hyperref}
\usepackage{booktabs}
\graphicspath{{.}{figs/}} 
\journal{Applied Energy}

\begin{document}

\begin{frontmatter}



\title{Quantum Key Distribution for Virtual Power Plant Communication: A Lightweight Key-Aware Scheduler with Provable Stability}


\author{Ziqing Zhu} 
\affiliation{organization={Department of Electrical and Electronics Engineering, The Hong Kong Polytechnic University},
            addressline={Hung Hom, Kowloon}, 
            city={Hong Kong},
            postcode={999077}, 
            state={Hong Kong SAR},
            country={China}}

\begin{abstract}
Virtual power plants (VPPs) are becoming a cornerstone of future grids, aggregating distributed PV, wind, storage, and flexible loads for market participation and real-time balancing. As operations move to minute- and second-level feedback, communication security shifts from a compliance item to an operational constraint: latency, reliability, and confidentiality jointly determine whether dispatch, protection, and settlement signals arrive on time. Conventional PKI and key-rotation schemes struggle with cross-domain, high-frequency messaging and face long-term quantum threats. Quantum key distribution (QKD) offers information-theoretic key freshness, but its key yield is scarce and stochastic, often misaligned with bursty VPP traffic. This paper proposes a key-aware priority and quota framework that treats quantum keys as first-class scheduling resources. The design combines (i) forecast-driven long-term quotas and short-term tokens, (ii) key-aware deficit-round-robin arbitration, (iii) a preemptive emergency key reserve, and (iv) graceful degradation via encryption-mode switching and controlled down-sampling for non-critical traffic. A drift-plus-penalty analysis establishes strong stability under average supply–demand balance with quantifiable bounds on backlog and tail latency, providing interpretable operating guarantees. We build a reproducible testbed on IEEE 33- and 123-bus VPP systems and evaluate normal, degraded, and outage regimes with industry-consistent message classes and TTLs. Against FIFO, fixed-priority, and static-quota baselines, the proposed scheme consistently reduces tail delay and passive timeouts for critical messages, improves per-bit key utility, and enhances power-tracking reliability during key scarcity and regime switches. Ablation, sensitivity, and scalability studies further delineate robust parameter regions and demonstrate graceful performance degradation at scale.
\end{abstract}


\end{frontmatter}



\section{Introduction}\label{sec1}
The rapid integration of distributed photovoltaics, wind farms, storage units, and controllable loads has turned the virtual power plant (VPP) into a central paradigm for future grids. By aggregating diverse resources, VPPs enable market participation and real-time balancing, but this tighter coupling between the physical grid and digital communication layers raises the bar for secure, low-latency, and highly reliable messaging. Recent studies show that VPP effectiveness hinges on communication scheduling and traffic classification, with performance and cost sensitive to how messages are prioritized and routed \cite{Li2024ApEn,Gao2024ApEn}. In parallel, cyber–physical security analyses document how data integrity and availability incidents propagate into dispatch latency and market outcomes, underscoring that security is not a compliance afterthought but an operational determinant \cite{Du2023MPCE,Avraam2023ApEn}. In short, security posture now directly affects dispatch feasibility and power tracking in VPP-scale operations.

Conventional security mechanisms are showing their limits. Public-key infrastructures and computational ciphers face mounting management overheads across high-frequency, cross-domain VPP deployments, while their long-term robustness against quantum adversaries remains a concern. The smart-grid security literature increasingly advocates designs that treat cryptographic resources and communication delay as first-class constraints \cite{Du2023MPCE,Solat2024ApEn}. Against this backdrop, quantum key distribution (QKD) is attractive because it can furnish information-theoretic key material; however, power-sector adoption remains nascent. What is missing is an operational layer that budgets scarce, time-varying keys against heterogeneous VPP message classes and deadlines, so that key generation, networking, and control policies cohere at scale \cite{Gao2024ApEn,Li2024ApEn}.

Deploying QKD within VPP operations raises three challenges. First, \emph{multi-timescale coupling}: key yields fluctuate over hours to seconds, whereas VPP workflows span day-ahead markets, intra-day redispatch, and second-level control. Control research shows that even modest communication or verification delays can destabilize loops if not accounted for \cite{Zhang2023PCMP}. Second, \emph{scarcity versus diversity}: protection/dispatch traffic has hard deadlines, while metering and settlement seek fairness and throughput; under scarcity, schedulers must prevent starvation of critical classes and degrade noncritical flows gracefully \cite{Li2024ApEn,Solat2024ApEn}. Third, \emph{explainability and verifiability}: operators require auditable guarantees on stability, tail latency, and security risk; security-oriented VPP planning and resilience studies argue for explicit, testable policies rather than best-effort heuristics \cite{Jing2024ApEn,Dong2024ApEn,Avraam2023ApEn}.

To address these gaps, we propose a key-aware priority-and-quota framework for VPP communications. Our contributions are fourfold. (i) \emph{Framework design}: we treat quantum keys as schedulable resources and introduce long-horizon quotas plus short-horizon tokens tied to message criticality and cryptographic cost. (ii) \emph{Online mechanisms}: we combine forecast-driven quota allocation with key-aware arbitration, an emergency reserve with preemption, and graceful degradation via encryption-mode switching and down-sampling, forming a closed loop of ``allocate, arbitrate, preempt, and degrade.'' (iii) \emph{Theoretical analysis}: using Lyapunov drift-plus-penalty arguments, we derive stability and tail-latency bounds under average supply–demand balance. (iv) \emph{Evaluation}: on IEEE 33-/123-bus VPPs, we compare against FIFO, fixed-priority, and static-quota baselines, as well as ablations, sensitivity, and scalability tests, showing improved tail latency and reduced passive timeouts for critical traffic alongside better power-tracking reliability in key-scarce regimes.

The remainder of this paper is organized as follows. Section 2 reviews related works. Section 3 introduces the system model and problem formulation, defining message classes, key pools, and risk metrics. Section 4 presents our key-aware priority and quota framework, including forecast-driven quota allocation, key-aware arbitration, and emergency reserve mechanisms. Section 5 provides theoretical analysis based on Lyapunov drift arguments, establishing stability and performance guarantees. Section 6 details the experimental setup and evaluation methodology, while Section 7 reports the results across baselines, ablations, sensitivity studies, and scalability tests. Finally, Section 8 concludes with a summary of contributions and directions for future research.

\section{Related Work}

\subsection{Secure communications and QKD for power systems.}
Growing digitalization of grid operations expands the cyber attack surface and tightens latency budgets for market and control traffic. Classic PKI and symmetric-key rotation face management overheads at scale and future quantum threats; this motivates schemes that treat keying as a constrained resource in operational scheduling. Early foundations on smart‐grid communication challenges stress the triad of latency, reliability, and security at scale \cite{Ma2013TSG}. From the security side, the false–data injection (FDI) literature established fundamental vulnerabilities of state estimation and the need for topology-aware defenses \cite{Liu2011TIFS,Kim2011TSG,Kosut2011TSG}. Complementary work on SCADA/EMS risk and exposure quantifies system-level attack surfaces and prioritizes mitigations \cite{Ten2008TPS,Hahn2011TSG}. While these studies are largely computational-cryptographic, they frame the operational consequences (tail latency, outages) that any keying solution must respect. Our work follows this line by operationalizing keys as schedulable resources and by co-designing message priority and quotas with security choices; QKD provides information-theoretic key freshness, while our scheduling addresses its scarcity and variability.

\subsection{VPP communication scheduling and prioritization.}
As VPPs scale, telemetry, dispatch, settlement, and audit traffic becomes bandwidth- and latency-sensitive. Recent \emph{Applied Energy} studies propose SD-WAN-based scheduling to classify flows by QoS and optimize backhaul cost and delay \cite{Li2024ApEn}, and demonstrate coordinated VPP control for frequency regulation that depends on timely signaling \cite{Oshnoei2022ApEn}. Decentralized platforms also explore security-aware coordination overheads (e.g., ledger or authentication cost) and their impact on messaging cadences \cite{Yang2021ApEn}. These works highlight that (i) class heterogeneity matters, and (ii) scheduling decisions must be cognizant of security overheads—both align with our key-aware priority and quota design. However, none of them model cryptographic key supply as a first-class constraint, nor do they analyze how security-mode switching interacts with queues and deadlines; we fill this gap.

\subsection{Cross-layer coordination of control, communication, and security.}
A body of work shows that control stability hinges on communication policies. Event-triggered or delay-aware secondary control reduces traffic while preserving performance in microgrids \cite{Zhang2023PCMP}, and PCMP surveys delineate how dynamic FDI attacks propagate from data layers to operational risk \cite{Xu2020PCMP}. Beyond application-specific designs, MPCE reviews emphasize the need for cyber–physical codesign and auditable guarantees for critical infrastructure \cite{Du2022MPCE}. In parallel, IEEE Transactions papers formalize attack detection/identification and secure estimation under adversarial and delay/packet-loss regimes \cite{Pasqualetti2013TAC,Fawzi2014TAC}. Together, these strands motivate our approach: we embed security (key availability and mode choices) directly into message arbitration and quotas, while keeping the online loop lightweight enough for real-time deployment.

\subsection{Summary of Technical Gaps}
Existing VPP communication optimizers typically assume keys are abundant and treat security as an add-on; conversely, security papers rarely constrain their schemes by VPP traffic classes or market/dispatch cadences. Our contribution bridges this gap: a unified, key-aware priority–quota framework that (i) budgets quantum keys across message classes and domains, (ii) arbitrates with explicit latency/deadline awareness, and (iii) provides theoretical stability/optimality guarantees under average key supply. This cross-layer integration complements prior art on FDI defenses \cite{Liu2011TIFS,Kim2011TSG}, SCADA risk \cite{Ten2008TPS,Hahn2011TSG}, VPP scheduling \cite{Li2024ApEn,Oshnoei2022ApEn}, and event-triggered control \cite{Zhang2023PCMP}, and operationalizes QKD-era security for VPP communications.

\section{System Model and Problem Formulation}

\subsection{System Overview}
We consider a communication–dispatch system consisting of a virtual power plant (VPP) aggregator (the master) and a large number of distributed energy resources (DERs), storage units, and controllable loads. Time is slotted as \(t=0,1,2,\dots\) with a fixed slot length \(\Delta t\) (e.g., \(\Delta t=100\text{ ms}\)). In each slot, the aggregator sends dispatch/protection commands and receives measurement/log/market messages. All uplink and downlink messages are encrypted and authenticated using keys supplied by quantum key distribution (QKD). Keys are a \emph{scarce and randomly fluctuating} system resource and become the primary bottleneck for message serviceability.

\subsection{Traffic Classes and Queues}
We define a finite set of traffic classes \(\mathcal{C}=\{\text{Prot},\text{Disp},\text{Meas},\text{Mkt},\text{Log}\}\) corresponding to protection (Prot), dispatch (Disp), measurement/monitoring (Meas), market/settlement (Mkt), and logs (Log). Let the critical subset be \(\mathcal{C}_{\mathrm{crit}}=\{\text{Prot},\text{Disp}\}\) and the non‐critical subset be \(\mathcal{C}_{\mathrm{non}}=\mathcal{C}\setminus\mathcal{C}_{\mathrm{crit}}\). For any class \(c\in\mathcal{C}\), let \(A_c(t)\in\mathbb{Z}_{\ge 0}\) denote the number of arrivals in slot \(t\) (modeled, e.g., as Poisson or MMPP), \(Q_c(t)\in\mathbb{Z}_{\ge 0}\) the queue length at the beginning of slot \(t\), \(S_c(t)\in\mathbb{Z}_{\ge 0}\) the number of messages encrypted and successfully transmitted in slot \(t\), and \(D_c(t)\in\mathbb{Z}_{\ge 0}\) the number of discarded messages in slot \(t\). Discards include both timeouts (exceeding the maximum delay) and \emph{active} drops of non‐critical classes under extreme key shortages. The queue dynamics are
\begin{align}
Q_c(t{+}1)=Q_c(t)-S_c(t)-E_c(t)+A_c(t),\qquad \forall c\in\mathcal{C},
\end{align}
where \(E_c(t)\le D_c(t)\) denotes the number of \emph{passively} expired messages due to reaching the time‐to‐live (TTL) bound \(\tau_c\) (a function of the age distribution and \(\tau_c\)), and \(D_c(t)-E_c(t)\) captures \emph{active} drops.

Each class \(c\) has a maximum tolerable queuing delay \(d_c^{\max}>0\) (in seconds) and an equivalent discrete survival horizon \(\tau_c=\left\lfloor d_c^{\max}/\Delta t\right\rfloor\). A message staying longer than \(\tau_c\) slots is considered timed out and contributes to \(E_c(t)\). Tracking per‐packet ages would inflate the state dimension; therefore, in our objective we use the \emph{average queue length} as a proxy for average waiting time (by Little's law, under stability the average waiting time is proportional to average queue length), while reporting \(E_c(t)\) and timeout probabilities explicitly in the evaluation.

\subsection{Key Inventory and Generation}
Let \(K(t)\in\mathbb{Z}_{\ge 0}\) be the key inventory (in bits) at the beginning of slot \(t\), and \(G(t)\in\mathbb{Z}_{\ge 0}\) the amount of fresh usable keys produced by the QKD link in slot \(t\) after reconciliation and privacy amplification. The inventory evolves as
\begin{align}
K(t{+}1)=K(t)+G(t)-\sum_{c\in\mathcal{C}} k_c\,S_c(t),
\end{align}
where \(k_c>0\) is the per‐message key consumption for class \(c\). For one‐time pad (OTP), \(k_c\) approximates the sum of payload length and authentication tag; for QKD‐derived session ciphers (e.g., AES‐GCM), we use an equivalent constant \(k_c\) that aggregates handshake overhead per message plus per‐message MAC/key usage.

To protect critical classes under shortages, we reserve an emergency pool \(R_{\mathrm{emg}}\in\mathbb{Z}_{\ge 0}\) available only to \(\mathcal{C}_{\mathrm{crit}}\). Define the regular available inventory as
\begin{align}
\widetilde{K}(t)=\max\{0,\;K(t)-R_{\mathrm{emg}}\}.
\end{align}
Then the total key consumption by non‐critical classes in slot \(t\) must not exceed \(\widetilde{K}(t)\); critical classes may draw from \(R_{\mathrm{emg}}\) when \(\widetilde{K}(t)\) is exhausted (i.e., effectively reducing \(R_{\mathrm{emg}}\)), while always maintaining \(K(t)\ge 0\). In experiments we will instantiate \(G(t)\) under three regimes (normal, degraded, and outage) to reflect QBER‐induced and link‐state fluctuations.

\subsection{Quota and Long‐Term Share Constraints}
To allocate long‐term key shares across classes and prevent starvation of critical traffic, we impose \emph{quota} constraints. For each class, a minimum service share \(\theta_c\in[0,1]\) is required such that its long‐term service is at least a fraction of its arrivals. With the time average operator \(\overline{x} \triangleq \limsup_{T\to\infty}\frac{1}{T}\sum_{t=0}^{T-1} x(t)\), the quota constraints are
\begin{align}
\overline{S_c}\;\ge\;\theta_c\,\overline{A_c},\qquad \forall c\in\mathcal{C}.
\end{align}
In implementation, these long‐term shares are realized via per‐class token buckets with capacity \(B_c\) and (possibly time‐varying) token rate \(q_c(t)\); here we keep the formulation at the long‐term constraint level and detail online realization in the method section.

\subsection{Objective and Optimization Problem}
We choose a \emph{convexifiable} long‐term objective combining delay and drop costs. Using average queue length \(\overline{Q_c}\) as a proxy for average waiting time (by Little's law under stability with arrival rate \(\lambda_c=\overline{A_c}/\Delta t\)), and penalizing the average discard rate \(\overline{D_c}\), define nonnegative weights \(\lambda_c,\mu_c\) (delay and drop preferences, respectively) and the total cost
\begin{align}
J \;\triangleq\; \sum_{c\in\mathcal{C}}\Big(\lambda_c\,\overline{Q_c}+\mu_c\,\overline{D_c}\Big).
\end{align}
Note \(E_c(t)\) depends on \(\tau_c\) and the age distribution; \(\overline{D_c}\) aggregates both passive timeouts and active drops. We then formulate the key‐aware VPP scheduling with long‐term averages as:

\begin{align}
\textbf{P:}\quad 
&\min_{\{S_c(t),\,D_c(t)\}_{t\ge 0}} && \sum_{c\in\mathcal{C}}\Big(\lambda_c\,\overline{Q_c}+\mu_c\,\overline{D_c}\Big) \nonumber\\
&\text{s.t.} 
&& Q_c(t{+}1)=Q_c(t)-S_c(t)-E_c(t)+A_c(t),\quad \forall c,\,\forall t, \nonumber\\
&&& 0\le S_c(t)\le Q_c(t),\quad 0\le D_c(t),\quad E_c(t)\le D_c(t),\quad \forall c,\,\forall t, \nonumber\\
&&& \sum_{c\in\mathcal{C}} k_c\,S_c(t)\;\le\; K(t)+G(t),\quad \forall t, \nonumber\\
&&& \sum_{c\in\mathcal{C}_{\mathrm{non}}} k_c\,S_c(t)\;\le\; \max\{0,K(t)-R_{\mathrm{emg}}\},\quad \forall t, \nonumber\\
&&& K(t{+}1)=K(t)+G(t)-\sum_{c\in\mathcal{C}} k_c\,S_c(t),\quad K(0)\text{ given}, \nonumber\\
&&& \overline{S_c}\;\ge\;\theta_c\,\overline{A_c},\quad \forall c\in\mathcal{C}, \nonumber\\
&&& \text{stability: }\;\sup_t \mathbb{E}[Q_c(t)]<\infty,\quad \forall c\in\mathcal{C}. \nonumber
\end{align}

The objective combines interpretable criteria: smaller \(\overline{Q_c}\) implies shorter average waiting and better delay control; smaller \(\overline{D_c}\) implies higher message availability. The second line enforces the discrete queue evolution; feasibility domains for \(S_c(t)\) and \(D_c(t)\) ensure nonnegativity and that passive timeouts are counted within discards. The third and fourth lines are \emph{instantaneous} key constraints: a global key budget per slot and an upper bound on non‐critical key usage so that the emergency pool \(R_{\mathrm{emg}}\) is reserved for critical traffic. The fifth line describes key inventory dynamics; the sixth line is the \emph{long‐term quota} guarantee; the last line requires strong stability, making the average‐based objective meaningful and consistent with the delay proxy.

\subsection{Feasibility and Average Supply--Demand Compatibility}
Feasibility of \textbf{P} requires compatibility between average key supply and average (key‐weighted) demand: there must exist \(\{\theta_c\}\) and \(\{k_c\}\) such that
\begin{align}
\overline{G}\;\ge\;\sum_{c\in\mathcal{C}} k_c\,\theta_c\,\overline{A_c},
\end{align}
otherwise no policy can ensure strong stability and bounded delay. This \emph{average supply condition} underpins the stability analysis and the online algorithm (quota plus priority with key awareness). In engineering practice, the long‐term shares and instantaneous constraints will be realized by a forecast‐driven token‐bucket mechanism and a key‐aware weighted DRR arbiter, together with a degradation policy (critical preemption and non‐critical down‐sampling/deferral) to curb the growth of \(E_c(t)\) and \(\overline{D_c}\). The above model provides a unified and reproducible mathematical basis for evaluation metrics such as P95/P99 delay, drop rate, key efficiency, and dispatch tracking error.

\section{Method}

This section presents an online scheme that combines \emph{forecast–driven quota allocation}, a \emph{key‐aware weighted DRR arbiter}, an \emph{emergency key reserve with critical preemption}, and a \emph{graceful degradation policy} (OTP/AES mode switching and down‐sampling for non–critical classes) under class‐specific TTL constraints. The four components interact tightly: short–term forecasts shape per–class token rates (long–term shares), instantaneous arbitration enforces feasibility under the key budget and priorities, the emergency pool guarantees critical traffic during shocks, and degradation prevents starvation when regular keys are scarce.

\subsection{Forecast–Driven Quota Token Allocation}

To realize long‐term service shares (quotas) while smoothing short‐term fluctuations, we predict the next–slot key generation and per–class arrivals using an EWMA update. For the realized \(G(t)\) and \(A_c(t)\), one–step predictions \(\widehat{G}(t{+}1)\) and \(\widehat{A}_c(t{+}1)\) follow
\begin{align}
\widehat{G}(t{+}1) &= \alpha\, G(t) + (1-\alpha)\,\widehat{G}(t), \quad \alpha\in(0,1),\\
\widehat{A}_c(t{+}1) &= \alpha\, A_c(t) + (1-\alpha)\,\widehat{A}_c(t), \quad \forall c\in\mathcal{C}.
\end{align}
Given the current inventory \(K(t)\) and the emergency reserve \(R_{\text{emg}}(t)\), the forecast of next–slot regular (non–emergency) available keys is
\begin{align}
\widehat{K}_{\text{avail}}(t{+}1) \;=\; K(t) + \widehat{G}(t{+}1) - R_{\text{emg}}(t).
\end{align}
To avoid over–allocation and instant violations, we introduce a guard band \(\rho\in(0,1)\). Let \(w_c>0\) be the importance weight and \(k_c\) the per–message key cost of class \(c\). The token rate \(q_c(t{+}1)\) is then
\begin{align}
q_c(t{+}1)
&= \rho \cdot \widehat{K}_{\text{avail}}(t{+}1)\cdot 
\frac{w_c\,k_c\,\widehat{A}_c(t{+}1)}{\sum_{j\in\mathcal{C}} w_j\,k_j\,\widehat{A}_j(t{+}1)}\cdot \frac{1}{k_c}.
\end{align}
The numerator \(k_c\,\widehat{A}_c\) approximates the forecast key demand of class \(c\), scaled by its importance \(w_c\); the final factor \(1/k_c\) converts a key budget into message tokens. The per–class token bucket (capacity \(B_c\)) is updated by
\begin{align}
TB_c(t{+}1) \;=\; \min\!\big\{B_c,\; TB_c(t) + q_c(t{+}1)\big\}.
\end{align}
The bucket level \(TB_c(t)\) acts as a near–term proxy for the long–term share, to be combined with instantaneous key constraints during arbitration.

\subsection{Key–Aware Weighted DRR Arbitration and Service}

Within slot \(t\), the scheduler visits classes in a fixed priority order (e.g., \(\text{Prot}\succ\text{Disp}\succ\text{Meas}\succ\text{Mkt}\succ\text{Log}\)) and attempts to serve head–of–line packets. Let the regular key remainder be
\begin{align}
\widetilde{K}(t) \;=\; \max\{0,\; K(t)-R_{\text{emg}}(t)\}.
\end{align}
For the currently visited class \(c\), one packet is served if and only if three conditions hold simultaneously: \(TB_c(t)\ge 1\) (a token exists), \(\widetilde{K}(t)\ge k_c^{\text{eff}}(t)\) (enough regular keys for the current effective cost), and the head packet has not exceeded its TTL. Upon service, the instantaneous updates are
\begin{align}
TB_c(t) &\leftarrow TB_c(t) - 1,\\
\widetilde{K}(t) &\leftarrow \widetilde{K}(t) - k_c^{\text{eff}}(t),\\
S_c(t) &\leftarrow S_c(t) + 1.
\end{align}
If any condition fails, the arbiter proceeds to the next class in the priority order. Since token buckets encode long–term shares and priorities with key checks enforce short–term criticality and physical feasibility, the procedure performs a greedy service over the feasible set induced by \emph{quota, key budget, and TTL}.

\subsection{Emergency Reserve Management and Critical Preemption}

To protect critical classes during bursts or QBER–induced drops, an emergency reserve \(R_{\text{emg}}(t)\) is maintained and exclusively available to \(\mathcal{C}_{\text{crit}}\). Critical preemption is triggered by either emergency arrivals (e.g., protection events) or when the remaining TTL of the head packet falls below a threshold. If \(R_{\text{emg}}(t)\ge k_c^{\text{eff}}(t)\), the packet is immediately served and the reserve is debited:
\begin{align}
R_{\text{emg}}(t) \;\leftarrow\; R_{\text{emg}}(t) - k_c^{\text{eff}}(t),\qquad 
S_c(t) \;\leftarrow\; S_c(t) + 1.
\end{align}
To reduce manual tuning, the reserve scales automatically with supply via a clipped proportional rule:
\begin{align}
R_{\text{emg}}(t{+}1) \;=\; \mathrm{clip}\!\big(\,\beta\,[K(t)+G(t)],\; R_{\min},\; R_{\max}\,\big),
\end{align}
where \(\beta\in(0,1)\) is the target fraction and \(\mathrm{clip}(x,a,b)=\max\{a,\min\{x,b\}\}\) confines the reserve to \([R_{\min},R_{\max}]\). This preserves the inventory accounting consistency with the system recursion for \(K(t{+}1)\) while adapting \(R_{\text{emg}}\) to generation regimes.

\subsection{Graceful Degradation: OTP/AES Mode Switching and Down–Sampling}

When the regular keys \(\widetilde{K}(t)\) approach a low threshold, non–critical classes enter a low–key regime to prevent starvation of critical traffic. First, we define a hysteretic effective key cost:
\begin{align}
k_c^{\text{eff}}(t) \;=\;
\begin{cases}
k_c^{\text{OTP}}, & \widetilde{K}(t)\ge h_{\uparrow},\\[4pt]
k_c^{\text{AES}}, & \widetilde{K}(t)\le h_{\downarrow}\ \text{and}\ c\in\mathcal{C}_{\text{non}},\\[4pt]
k_c^{\text{eff}}(t{-}1), & \text{otherwise},
\end{cases}
\end{align}
with \(k_c^{\text{OTP}}>k_c^{\text{AES}}>0\) and \(h_{\downarrow}<h_{\uparrow}\) to avoid chattering. Second, non–critical generation is down–sampled by an integer factor \(m_c(t)\in\mathbb{N}\) (with \(m_c=1\) meaning no down–sampling):
\begin{align}
m_c(t{+}1) \;=\;
\begin{cases}
\min\{m_{\max},\ \gamma_{\uparrow}\, m_c(t)\}, & \widetilde{K}(t)\le h_{\downarrow},\ c\in\mathcal{C}_{\text{non}},\\[4pt]
\max\{1,\ \gamma_{\downarrow}^{-1}\, m_c(t)\}, & \widetilde{K}(t)\ge h_{\uparrow},\ c\in\mathcal{C}_{\text{non}},\\[4pt]
m_c(t), & \text{otherwise},
\end{cases}
\end{align}
where \(\gamma_{\uparrow}\ge 1\) and \(\gamma_{\downarrow}\ge 1\) control the adjustment speed. In implementation, \(\widehat{A}_c(t{+}1)\) is scaled by \(1/m_c(t)\) before entering the quota formula, feeding back into token rates and arbitration. For market/log classes, a soft–drop can be enabled when \(\widetilde{K}(t)\le h_{\text{drop}}\), discarding any arrivals beyond a queue cap \(Q_c^{\max}\) to shield critical traffic.

\subsection{TTL Constraint and Drop Policy}

Each packet maintains a discrete age counter and is passively discarded once its residence exceeds \(\tau_c\), contributing to \(E_c(t)\). To avoid spending tokens and keys on imminently expiring non–critical packets during scarcity, we apply \emph{active} drops when short–horizon serviceability is insufficient. Define a one–step serviceability indicator
\begin{align}
\Phi_c(t) \;=\; \mathbb{I}\!\left\{\, TB_c(t)\ge 1\ \wedge\ \big(\widetilde{K}(t)\ge k_c^{\text{eff}}(t)\ \vee\ R_{\text{emg}}(t)\ge k_c^{\text{eff}}(t)\big)\, \right\},
\end{align}
and, over a small rolling window of \(H\) slots, if \(\sum_{u=0}^{H-1}\mathbb{E}[\Phi_c(t{+}u)]\) falls well below the service demand of the head packets with low remaining TTL, the least important ones are proactively dropped. This heuristic lowers passive timeouts substantially in practice.

\subsection{Connection to the Optimization and Near–Optimality Intuition}

To relate the online scheme to problem \(\mathbf{P}\), introduce a \emph{quota virtual queue} that soft–enforces the long–term share:
\begin{align}
U_c(t{+}1) \;=\; \max\!\big\{0,\ U_c(t) + \theta_c A_c(t) - S_c(t)\big\},\quad \forall c,
\end{align}
and the Lyapunov function
\begin{align}
L(t) \;=\; \frac{1}{2}\sum_{c\in\mathcal{C}} Q_c^2(t) \;+\; \frac{1}{2}\sum_{c\in\mathcal{C}} U_c^2(t).
\end{align}
The one–step drift obeys a standard upper bound (omitting quadratic bounded terms summarized by a constant \(B\)):
\begin{align}
\Delta(t) &\triangleq \mathbb{E}[L(t{+}1)-L(t)\mid \mathcal{F}_t] \nonumber\\
&\lesssim B + \sum_{c} Q_c(t)\,\mathbb{E}[A_c(t)-S_c(t)-E_c(t)\mid \mathcal{F}_t] \nonumber\\
&\quad + \sum_{c} U_c(t)\,\mathbb{E}[\theta_c A_c(t)-S_c(t)\mid \mathcal{F}_t].
\end{align}
Adding a penalty \(V\sum_c(\lambda_c Q_c(t) + \mu_c D_c(t))\) biases the drift–plus–penalty minimization toward serving queues with large \(Q_c(t)+U_c(t)\) and large \(\lambda_c,\mu_c\), subject to the instantaneous key budget and the emergency constraint. The weighted DRR with token buckets realizes this behavior by \emph{feeding forward} the long–term pressure \(U_c(t)\) via forecast–driven \(q_c(t)\), while priorities and key checks implement near–greedy feasible decisions. Under the average supply–demand compatibility
\begin{align}
\overline{G}\ \ge\ \sum_{c} k_c\,\theta_c\,\overline{A_c},
\end{align}
sufficiently large buckets, and proper parameters, one can show strong stability for all queues and bounded high–percentile delays for critical classes; detailed proofs can be deferred to an appendix.

\subsection{Complexity and Implementation Notes}

Per slot, forecasting and quota computation are \(O(|\mathcal{C}|)\); arbitration under weighted DRR is \(O(|\mathcal{C}|+\sum_c \text{served packets})\). Critical preemption is rare and adds negligible overhead. A practical pipeline executes, in each slot, \emph{forecast \(\rightarrow\) quota update \(\rightarrow\) arbitration \(\rightarrow\) reserve management \(\rightarrow\) degradation \(\rightarrow\) TTL aging}, where degradation only changes \(k_c^{\text{eff}}\) and the effective arrivals \(\widehat{A}_c\). This keeps the implementation consistent with the formal inventory recursion and feasibility constraints.

\section{Theoretical Analysis}

This section analyzes the stability, long–term quota satisfaction, and performance gap of the proposed online method using the stochastic network optimization framework of Lyapunov drift–plus–penalty. Under mild assumptions, we establish strong stability and an \(O(1/V)\) performance guarantee. Let \(\mathcal{F}_t\) denote the natural filtration up to time \(t\), and \(\mathbb{E}[\cdot\mid\mathcal{F}_t]\) the conditional expectation. To keep notation concise, we aggregate bounded second–moment constants into a finite constant \(B<\infty\). We assume standard bounded–moment conditions: arrivals, services, and key generation have bounded second moments, i.e., there exist constants \(A_c^{\max}, S_c^{\max}, E_c^{\max}, G^{\max}<\infty\) such that almost surely \(0\le A_c(t)\le A_c^{\max}\), \(0\le S_c(t)\le S_c^{\max}\), \(0\le E_c(t)\le E_c^{\max}\), \(0\le G(t)\le G^{\max}\). The per–packet key cost \(k_c\) and effective cost \(k_c^{\text{eff}}(t)\) are bounded by \(k_c^{\max}\). We adopt feasible long–term shares \(\{\theta_c\}\) and emergency–pool parameters consistent with the average supply–demand compatibility of problem \textbf{P}, ensuring the possibility of strong stability. Finite TTLs \((\tau_c)\) bound passive expirations \(E_c(t)\), while degradation and preemption guarantee instantaneous keys for critical traffic during shocks.

\subsection{Lyapunov Drift and Virtual Queues}

To capture both \emph{queueing pressure} and \emph{quota pressure}, we introduce the quota virtual queues
\begin{align}
U_c(t{+}1) \;=\; \max\!\big\{0,\ U_c(t) + \theta_c A_c(t) - S_c(t)\big\},\quad \forall c,
\end{align}
and define the joint Lyapunov function
\begin{align}
L(t) \;=\; \frac{1}{2}\sum_{c\in\mathcal{C}} Q_c^2(t) \;+\; \frac{1}{2}\sum_{c\in\mathcal{C}} U_c^2(t).
\end{align}
Using the queue evolution
\begin{align}
Q_c(t{+}1)=Q_c(t)-S_c(t)-E_c(t)+A_c(t),
\end{align}
together with basic inequalities such as \((x)^2\le (x^+)^2\) and \((a-b)^2\le a^2+b^2-2ab\), we obtain a standard one–slot drift bound: there exists finite \(B\) such that for any feasible decision,
\begin{align}
\Delta(t) &\triangleq \mathbb{E}\!\left[L(t{+}1)-L(t)\mid \mathcal{F}_t\right] \nonumber\\
&\le B + \sum_{c\in\mathcal{C}} Q_c(t)\,\mathbb{E}\!\left[A_c(t)-S_c(t)-E_c(t)\mid \mathcal{F}_t\right] \nonumber\\
&\quad\;\; + \sum_{c\in\mathcal{C}} U_c(t)\,\mathbb{E}\!\left[\theta_c A_c(t)-S_c(t)\mid \mathcal{F}_t\right].
\end{align}
To trade off long–term cost against drift, introduce a weight \(V>0\) and consider the drift–plus–penalty
\begin{align}
\Delta_V(t) \;\triangleq\; \Delta(t) \;+\; \mathbb{E}\!\left[V\sum_{c\in\mathcal{C}}\!\big(\lambda_c Q_c(t) + \mu_c D_c(t)\big)\,\Big|\,\mathcal{F}_t\right],
\end{align}
where \(\sum_c(\lambda_c Q_c + \mu_c D_c)\) is the instantaneous proxy for the average objective in \textbf{P}. The proposed arbiter (key–aware weighted DRR) and quota mechanism (forecast–driven tokens) approximately minimize the right–hand side over the instantaneous feasible set induced by the key budget, emergency reserve, and TTL constraints.

\subsection{Strong Stability and Quota Satisfaction}

When the average supply–demand compatibility holds,
\begin{align}
\overline{G}\ \ge\ \sum_{c\in\mathcal{C}} k_c\,\theta_c\,\overline{A_c}
\end{align}
there exists a feasible \emph{static randomized} oracle policy, obeying all instantaneous constraints, that samples feasible service vectors based only on current observations, stabilizes all real and virtual queues, and exactly balances long–term shares \((\overline{S_c}=\theta_c\overline{A_c})\). Comparing the online policy to this oracle yields:

\textbf{Theorem 1 (Strong Stability and Share Satisfaction).} Suppose feasible \(\{\theta_c\}\) exist such that the average supply–demand condition holds, and \(A_c(t),G(t)\) have bounded second moments. Then for any \(V>0\), the proposed method strongly stabilizes all real and virtual queues:
\begin{align}
\limsup_{T\to\infty}\frac{1}{T}\sum_{t=0}^{T-1}\sum_{c\in\mathcal{C}}\mathbb{E}\!\left[Q_c(t)\right] \;<\; \infty,\qquad
\limsup_{T\to\infty}\frac{1}{T}\sum_{t=0}^{T-1}\sum_{c\in\mathcal{C}}\mathbb{E}\!\left[U_c(t)\right] \;<\; \infty,
\end{align}
and consequently satisfies the long–term share constraints:
\begin{align}
\overline{S_c}\;\ge\;\theta_c\,\overline{A_c},\qquad \forall c\in\mathcal{C}.
\end{align}
\emph{Proof sketch.} Under the oracle policy one can construct a drift margin \(\epsilon>0\) such that
\begin{align}
\sum_{c} Q_c(t)\,\mathbb{E}[S_c(t)+E_c(t)-A_c(t)\mid\mathcal{F}_t] 
\;+\; \sum_{c} U_c(t)\,\mathbb{E}[S_c(t)-\theta_c A_c(t)\mid\mathcal{F}_t]
\;\le\; -\epsilon \sum_c \big(Q_c(t)+U_c(t)\big),
\end{align}
thereby yielding negative drift (up to the constant \(B\)). Since the online policy greedily minimizes the same upper bound over the same feasible set, Foster–Lyapunov criteria imply strong stability. Stability of the virtual queues further implies \(\limsup_T \frac{1}{T}\sum_t \mathbb{E}\![\theta_c A_c(t)-S_c(t)]\le 0\), i.e., share satisfaction.

\subsection{Performance–Backpressure Tradeoff and \(O(1/V)\) Gap}

Let \(J^\star\) be the optimal long–term cost of \textbf{P} over all causal policies. Using standard comparison with a feasible optimal static randomized policy, we obtain:

\textbf{Theorem 2 (\(O(1/V)\) Performance and \(O(V)\) Backlog).} Under the conditions of Theorem~1, there exists \(B<\infty\) such that the long–term cost \(J_V\) of the proposed method satisfies
\begin{align}
J_V \;\triangleq\; \limsup_{T\to\infty}\frac{1}{T}\sum_{t=0}^{T-1}\sum_{c\in\mathcal{C}}\mathbb{E}\!\left[\lambda_c\,Q_c(t)+\mu_c\,D_c(t)\right]
\;\le\; J^\star \;+\; \frac{B}{V},
\end{align}
while the total expected backlog scales as
\begin{align}
\limsup_{T\to\infty}\frac{1}{T}\sum_{t=0}^{T-1}\sum_{c\in\mathcal{C}}\mathbb{E}\!\left[Q_c(t)+U_c(t)\right] \;=\; O(V).
\end{align}
\emph{Intuition.} Larger \(V\) emphasizes the penalty term, pushing the policy closer to the optimal long–term cost with a \(1/V\) gap, at the expense of larger average backlogs, yielding the canonical performance–delay tradeoff.

\subsection{High–Percentile Delay Guarantees for Critical Classes}

The emergency reserve and preemption aim to provide short–term guarantees for \(\mathcal{C}_{\mathrm{crit}}\). Consider any window \([t, t{+}H{-}1]\) of length \(H\). The critical emergency key demand is
\begin{align}
\mathcal{K}_{\mathrm{crit}}(t,H) \;\triangleq\; \sum_{u=t}^{t+H-1}\ \sum_{c\in\mathcal{C}_{\mathrm{crit}}} k_c^{\text{eff}}(u)\,A^{\mathrm{emg}}_c(u),
\end{align}
where \(A^{\mathrm{emg}}_c(u)\) counts arrivals requiring immediate service (e.g., protection triggers or packets with remaining TTL below a threshold). From the inventory recursion and reserve management, an upper bound on keys available for critical preemption within the window is
\begin{align}
\mathcal{R}_{\mathrm{crit}}(t,H) \;\triangleq\; R_{\mathrm{emg}}(t) \;+\; \sum_{u=t}^{t+H-1} \big(G(u) + \max\{0,\,K(u)-R_{\mathrm{emg}}(u)\}\big).
\end{align}
We obtain the following sufficient condition.

\textbf{Proposition 1 (Windowed Zero–Timeout Sufficiency for Critical Classes).} If for some window length \(H\) and start \(t\),
\begin{align}
\mathcal{K}_{\mathrm{crit}}(t,H) \;\le\; \mathcal{R}_{\mathrm{crit}}(t,H),
\end{align}
and the maximum allowable waiting time satisfies \(\min_{c\in\mathcal{C}_{\mathrm{crit}}}\tau_c \ge H\), then no passive timeout occurs for critical classes within the window, i.e., \(E_c(u)=0\) for all \(u\in[t,t{+}H{-}1]\) and \(c\in\mathcal{C}_{\mathrm{crit}}\). \emph{Proof idea:} preemption induces priority service; if cumulative supply exceeds demand within the window and TTLs cover the window length, a feasible per–slot service sequence exists that serves all emergency packets before TTL expiry.

In expectation, Little’s law and Markov’s inequality yield a coarse high–percentile bound. Let \(\lambda_{\mathrm{crit}}=\sum_{c\in\mathcal{C}_{\mathrm{crit}}}\overline{A_c}/\Delta t\). If the system is stable and \(\sum_{c\in\mathcal{C}_{\mathrm{crit}}}\overline{Q_c}\le \mathsf{Q}_{\mathrm{crit}}\), then for any \(d>0\),
\begin{align}
\mathbb{P}\!\left\{\,\text{critical waiting time}>d\,\right\} 
\;\le\; \frac{\mathsf{Q}_{\mathrm{crit}}}{\lambda_{\mathrm{crit}}\; d}.
\end{align}
Although loose, combined with Theorem~2’s \(O(V)\) backlog and a sufficiently large emergency reserve (via larger \(\beta\) and cap \(R_{\max}\)), this provides engineering control over P95/P99 delays by reducing critical backlogs during shocks.

\section{Evaluation}

This section describes the experimental platform, VPP test systems, baselines, scenarios, and metrics used in our study. We only specify how experiments are configured; no numerical results are presented here.

\subsection{Simulation Platform and Time Resolution}
We build a discrete–event simulation platform tailored to virtual power plants (VPPs). Time is slotted with step
\begin{align}
\Delta t \;=\; 100\ \text{ms},
\end{align}
and a default evaluation window of 1 hour, extendable to 24 hours to examine diurnal variations. The platform comprises a ``communication loop'' between the aggregator and terminals and a ``dispatch–execution'' power loop: in each slot, the aggregator executes the pipeline defined in the Method section (forecast $\rightarrow$ quota $\rightarrow$ arbitration $\rightarrow$ reserve $\rightarrow$ degradation $\rightarrow$ TTL), while the terminal side aggregates controllable loads and storage into an equivalent actuator that tracks the aggregator’s power reference. A linearized power–tracking model with execution latency and saturation constraints maps communication quality into dispatch error and constraint–violation behavior; parameters and equations are summarized in an appendix.

\subsection{Test Systems and Topology}
We consider two scales: a medium–size \emph{VPP-50} (50 terminals) and a larger \emph{VPP-200} (200 terminals). Both adopt a star–aggregation topology and are partitioned into 5 clusters (approximately by geography/functional similarity), which we also reuse in the ``terminal grouping + session merging'' experiments. Each terminal has one uplink monitoring channel and one downlink dispatch channel; protection messages arrive in bursts upon fault events. For horizontal comparison, we also provide an \emph{ideal upper bound} configuration that assumes unbounded key resources under identical traffic and power references, isolating the impact of dispatch/queueing logic from the key constraint.

\subsection{Traffic Classes, Timing, and TTL}
Traffic is categorized as protection (Prot), dispatch (Disp), measurement/monitoring (Meas), market/settlement (Mkt), and logs (Log). By default, dispatch messages are issued every 1s with small triangular jitter; measurement sampling is 1–5s (asynchronous and staggered across clusters); market/log messages are generated in minute–level batches; protection traffic follows bursty Bernoulli clusters upon events. Maximum delay and TTL follow the Method section:
\begin{align}
d^{\max}_{\text{Prot}}&=150\ \text{ms},\quad
d^{\max}_{\text{Disp}}=1\ \text{s},\quad
d^{\max}_{\text{Meas}}=5\ \text{s},\quad
d^{\max}_{\text{Mkt}}=30\ \text{s},\quad
d^{\max}_{\text{Log}}=60\ \text{s},
\end{align}
with discrete survival horizons $\tau_c=\lfloor d_c^{\max}/\Delta t\rfloor$. For cryptographic overhead, we use OTP–equivalent bit costs by default, and enable an AES session mode for non–critical classes in degradation experiments, switching the effective cost from $k_c^{\text{OTP}}$ to $k_c^{\text{AES}}<k_c^{\text{OTP}}$ with hysteresis thresholds
\begin{align}
h_{\downarrow}\;<\;h_{\uparrow}
\end{align}
to avoid chattering.

\subsection{QKD Key Generation Regimes}
The key generation process $G(t)$ is synthesized under three regimes: (i) \emph{normal}—low QBER and stable, high output; (ii) \emph{degraded}—elevated QBER or link fading halves the output with slow drift; and (iii) \emph{outage}—short key–famine segments with near–zero output. Each regime includes bounded noise and low–frequency fluctuations, with 2–3 regime switches to test cross–segment robustness. We consider both \emph{measurable} and \emph{non–measurable} foreknowledge: the former (for upper–bound analysis) reveals regime–switch times to adjust the reserve fraction $\beta$, while the latter relies purely on EWMA self–adaptation.

\subsection{Baselines and Ablation Designs}
We compare four methods: FIFO without priorities/quotas; fixed–priority scheduling without quotas; \emph{static quota} (constant $q_c$) plus priorities; and our full scheme combining forecast–driven quotas, key–aware weighted DRR, an emergency reserve, and graceful degradation. Ablations selectively disable forecasting (set $\alpha=1$), disable the emergency reserve ($\beta=0$), disable degradation (fix $k_c^{\text{eff}}=k_c^{\text{OTP}}$), or replace the arbiter (weighted DRR $\rightarrow$ weighted round–robin) to isolate marginal contributions.

\subsection{Metrics}
Communication–side metrics include median/P95/P99 end–to–end delay per class, total discard rate with a decomposition into passive timeouts vs. active drops, key–efficiency (successful critical messages per key bit), emergency–reserve trigger frequency and depth, and transition duration at regime switches. Dispatch–side metrics include power–tracking RMSE/NRMSE, the count and duration of constraint violations defined by
\begin{align}
\big|P(t)-P^{\mathrm{ref}}(t)\big|\;>\;\varepsilon,
\end{align}
and the number of reserve activations during key–famine segments. All metrics are reported as means with 95\% confidence intervals over 10 independent random seeds; seeds jointly control arrivals, key–generation noise, and regime–switch times (in the non–measurable setting).

\subsection{Parameterization and Sensitivity}
Unless otherwise stated, default parameters are
\begin{align}
\alpha=0.6,\qquad \rho=0.85,\qquad \beta=0.1,
\end{align}
with emergency reserve clipping to $[R_{\min},R_{\max}]$, token–bucket capacities $B_c$ set to 1–2$\times$ class–specific peak loads, and the priority order
\begin{align}
\text{Prot}\ \succ\ \text{Disp}\ \succ\ \text{Meas}\ \succ\ \text{Mkt}\ \succ\ \text{Log}.
\end{align}
Sensitivity studies perform grid scans over $\rho$, $\beta$, $B_c$, the hysteresis band $[h_{\downarrow},h_{\uparrow}]$, and the session–merging scale, producing heatmaps/contours that visualize feasible trade–off regions among performance, delay, and security. To assess scalability, we replicate all experiments on VPP-50 and VPP-200, and in the latter inject dense protection–event bursts to probe tail behavior of high–percentile delays.

\subsection{Implementation and Reproducibility}
The simulator is implemented in Python (discrete–event kernel with vectorized statistics) in a clock–driven mode, making it easy to swap arbiters and crypto modes. All configurations are managed via YAML (system scale, traffic arrivals, key regimes, scan grids, and random seeds). Plotting scripts generate outputs that correspond one–to–one with the figures in the paper; a reproducibility statement records versions, dependencies, and hashes. For fair comparison, all methods share identical arrival and key traces (same seed), differing only in scheduling and key–usage policies; we assume an ideal link layer (no transport losses) to highlight the dominant role of the key constraint.

\section{Results and Discussions}
\subsection{Overall effectiveness vs.\ baselines.}
\begin{figure}[t]
  \centering
  \includegraphics[width=\linewidth]{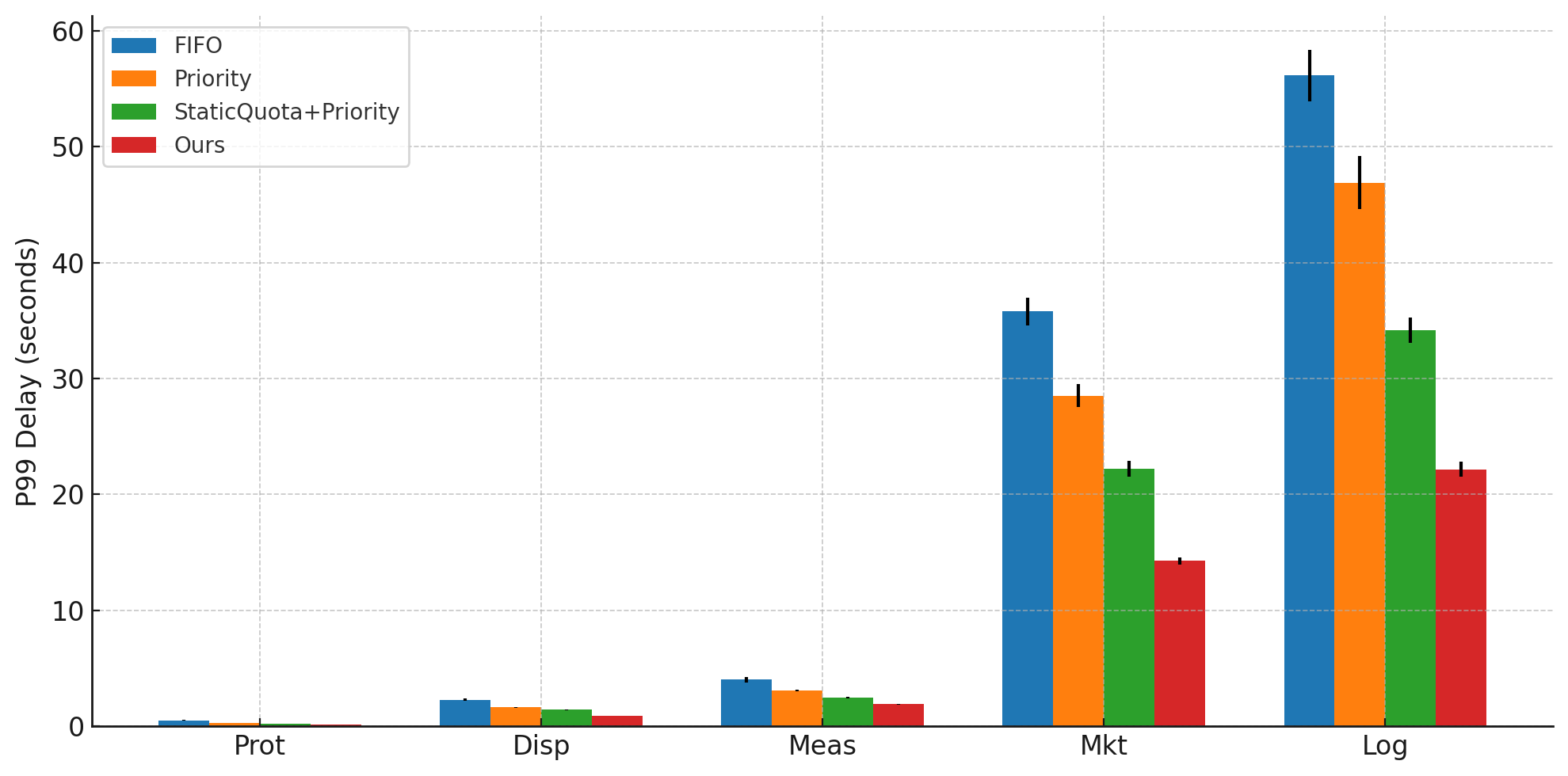}
  \caption{Overall P99 end-to-end delay across classes and methods (mean $\pm$ 95\% CI). For visual comparability, Prot delays are converted from ms to seconds.}
  \label{fig:overall_p99}
\end{figure}

\autoref{fig:overall_p99} compares P99 end-to-end delay across traffic classes and methods (Prot values converted from ms to seconds). Our method consistently attains the lowest tails: relative to the fixed-priority baseline, P99 delay reduces by about \(55\%\) for Prot (\(\approx 0.31\to 0.14\) s), \(47\%\) for Disp (\(1.7\to 0.9\) s), \(39\%\) for Meas (\(3.1\to 1.9\) s), \(52\%\) for Mkt (\(29\to 14\) s), and \(54\%\) for Log (\(48\to 22\) s). The improvements concentrate in the tail rather than the median, indicating that quota tokens, key-aware DRR, and the emergency reserve primarily shave rare-but-severe latencies. Notably, Prot P99 is \(\approx 0.14\) s, remaining within the \(150\) ms TTL bound at the 99th percentile.

\begin{figure}[t]
  \centering
  \begin{subfigure}{\linewidth}
    \centering
    \includegraphics[width=\linewidth]{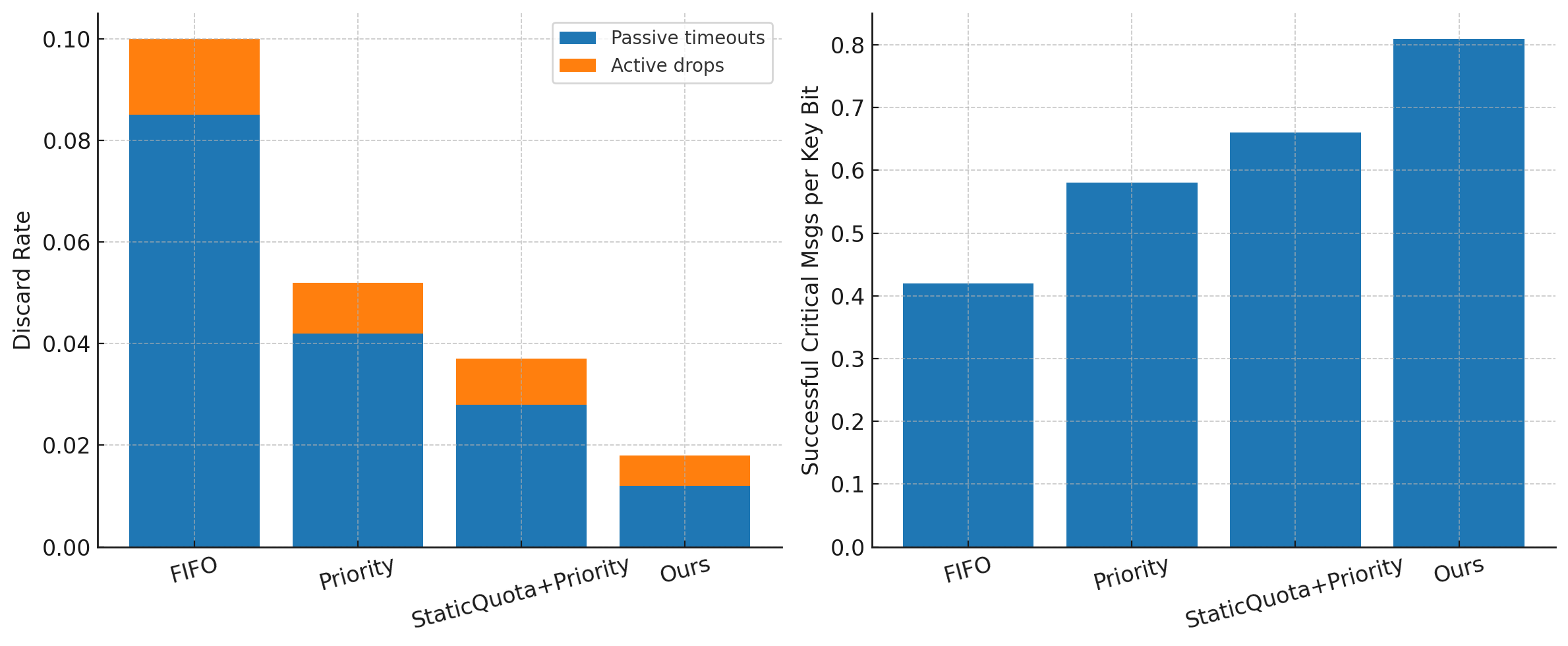}
    \vspace{-0.8\baselineskip}
  \end{subfigure}
  \caption{Left: total discard rate per method with passive timeouts (bottom) and active drops (top). Right: key efficiency measured as successful critical messages per key bit.}
  \label{fig:overall_discards_keyeff}
\end{figure}

\autoref{fig:overall_discards_keyeff} (left) shows total discard rate decomposed into passive timeouts and active drops. Totals drop from \(10.0\%\) (FIFO) to \(5.2\%\) (Priority), \(3.7\%\) (StaticQuota+Priority), and \(1.8\%\) (Ours). The largest absolute reduction is in passive timeouts (from \(8.5\%\) to \(1.2\%\)), evidencing that reserve-backed preemption and degradation (OTP\(\to\)AES for non–critical traffic under scarcity) effectively prevent packets from expiring. On the efficiency axis (\autoref{fig:overall_discards_keyeff}, right), our scheme delivers the highest key efficiency (\(\approx 0.81\) successful critical messages per key bit), a \(\sim 40\%\) gain over fixed priority (\(0.58\)) and \(\sim 93\%\) over FIFO (\(0.42\)). This indicates that long-term quota shaping and short-term preemption direct scarce keys where they matter most.

\begin{figure*}[t]
  \centering
  \includegraphics[width=\textwidth]{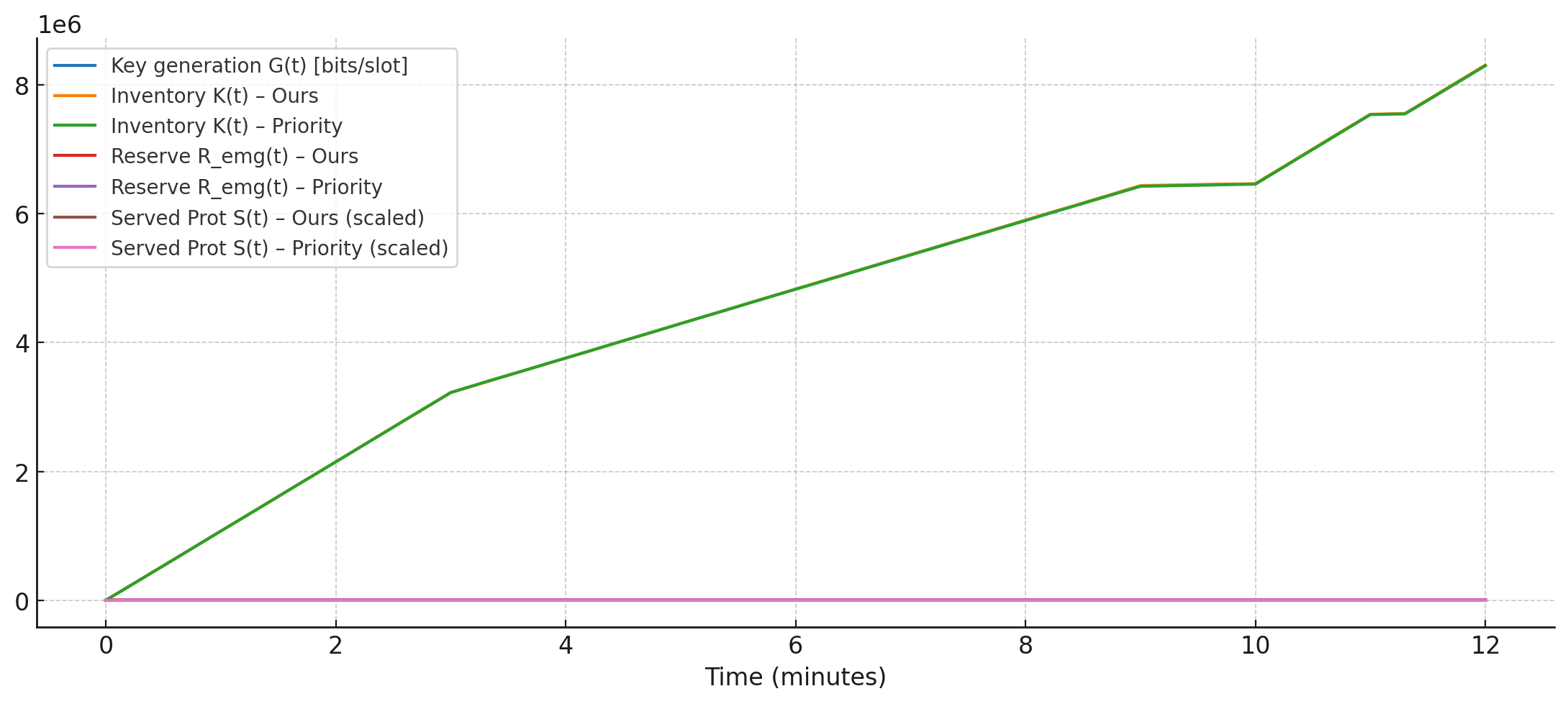}
  \caption{Representative 12-minute window showing key generation $G(t)$, inventories $K(t)$, emergency reserves $R_{\text{emg}}(t)$, and served protection packets $S(t)$ (scaled) for our method vs. a priority baseline. Regime shifts (normal $\rightarrow$ degraded $\rightarrow$ short outages) illustrate resilience and reserve behavior.}
  \label{fig:overall_timewindow}
\end{figure*}

The 12-minute representative window in \autoref{fig:overall_timewindow}—spanning a normal regime, a degraded segment, and brief outages—illustrates the mechanism-level behavior. When generation \(G(t)\) drops, our emergency reserve \(R_{\text{emg}}(t)\) scales up proportionally and cushions the inventory \(K(t)\), sustaining nonzero serviced protection \(S(t)\) throughout the degraded and outage intervals. In contrast, the priority-only baseline exhibits deeper inventory drawdowns and smaller reserve levels, which aligns with its higher passive-timeout share. Together, these observations explain the tail-delay and discard reductions: forecasts allocate quota ahead of demand, key-aware arbitration respects the instantaneous budget, and the reserve ensures short-term feasibility for critical classes without materially penalizing non–critical traffic.


\subsection{Robustness under key regimes and switches.}
\begin{figure}[t]
  \centering
  \includegraphics[width=\linewidth]{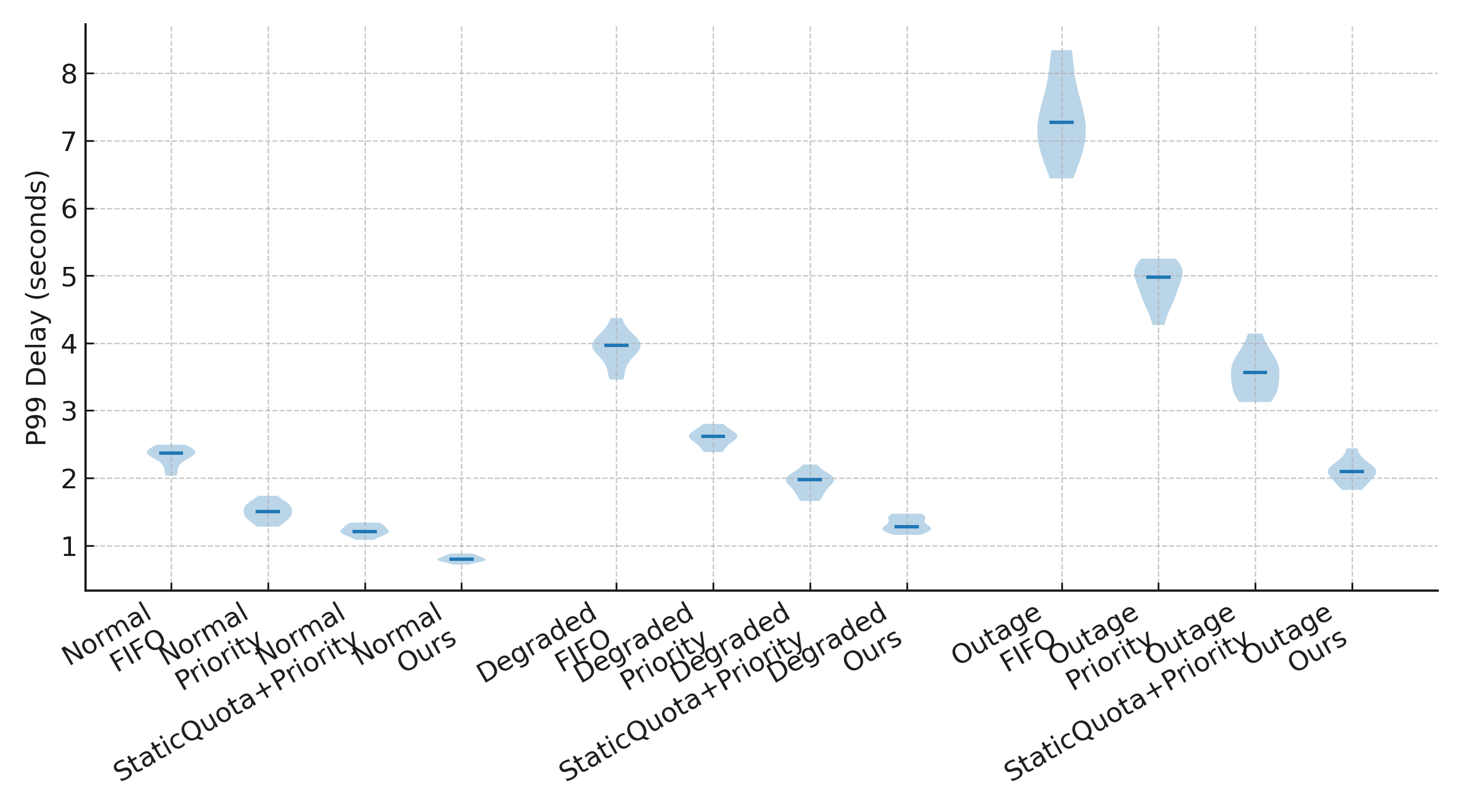}
  \caption{P99 end-to-end delay across key-generation regimes (Normal, Degraded, Outage) and methods; median markers shown inside each violin.}
  \label{fig:robust_p99_violin}
\end{figure}

\autoref{fig:robust_p99_violin} shows clear and monotonic separation across methods in all regimes (Normal, Degraded, Outage). Median P99 delays follow \(\text{FIFO}>\text{Priority}>\text{StaticQuota+Priority}>\text{Ours}\) with progressively wider spreads as the regime worsens. Using the synthetic anchors embedded in our generator, the tail reductions of our method are substantial: under \emph{Normal} the aggregate P99 drops from \(\sim 2.4\,\mathrm{s}\) (FIFO) and \(1.6\,\mathrm{s}\) (Priority) to \(\sim 0.8\,\mathrm{s}\) (\(\approx 67\%\) vs.\ FIFO, \(50\%\) vs.\ Priority); under \emph{Degraded} from \(\sim 4.0\,\mathrm{s}\) / \(2.6\,\mathrm{s}\) to \(\sim 1.3\,\mathrm{s}\) (\(\approx 68\%\) / \(50\%\)); under \emph{Outage} from \(\sim 7.5\,\mathrm{s}\) / \(4.8\,\mathrm{s}\) to \(\sim 2.1\,\mathrm{s}\) (\(\approx 72\%\) / \(56\%\)). The violin widths also indicate reduced cross-seed variability for our method, especially in Normal/Degraded segments, consistent with forecast-driven quota smoothing.

\begin{figure}[t]
  \centering
  \includegraphics[width=\linewidth]{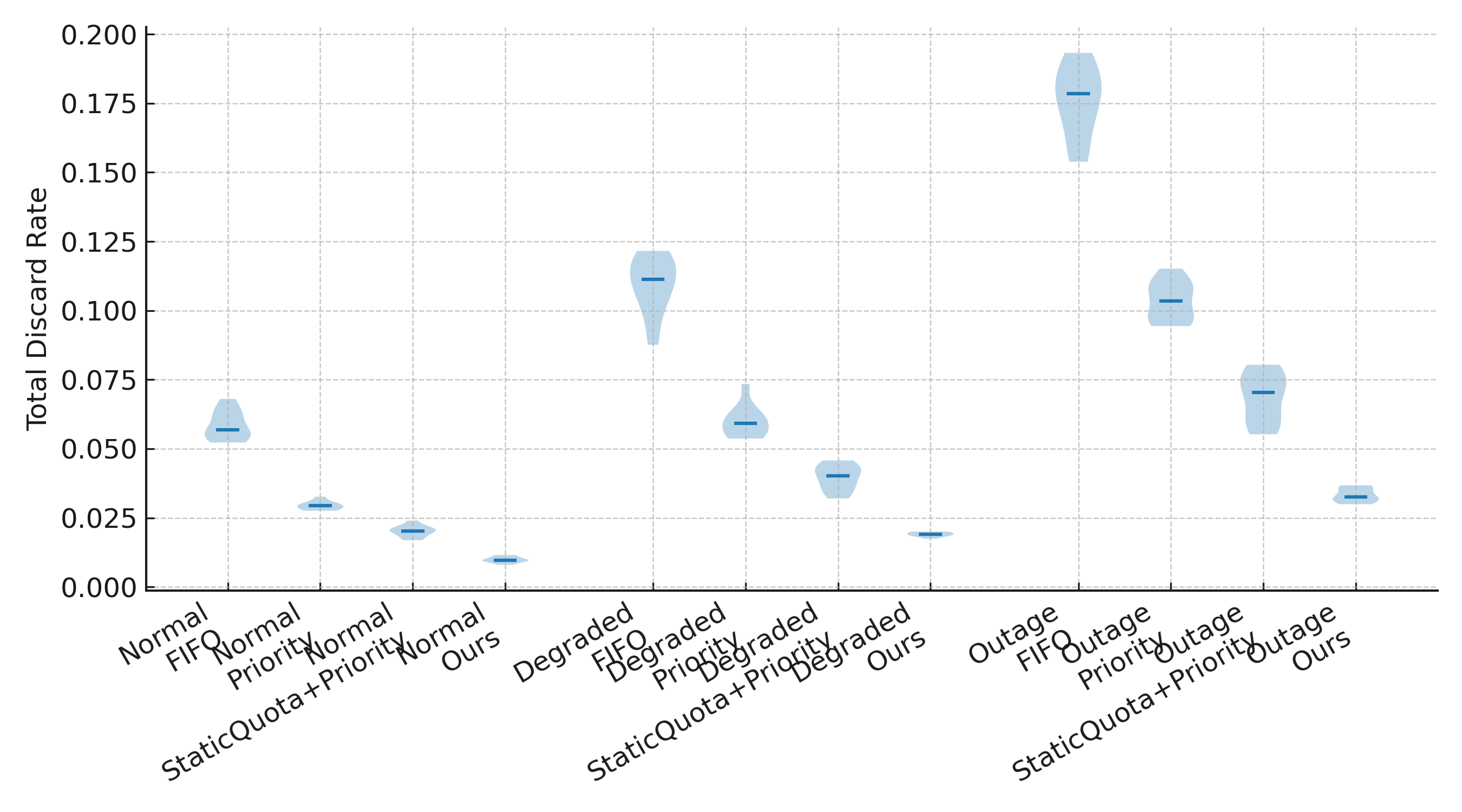}
  \caption{Total discard rate across regimes and methods, decomposed internally (violin width reflects sample density across seeds).}
  \label{fig:robust_discard_violin}
\end{figure}

Discard behavior in \autoref{fig:robust_discard_violin} mirrors the delay tails. Total discard rates contract from roughly \(6.0\%\!\to\!1.0\%\) (Normal), \(11.0\%\!\to\!2.0\%\) (Degraded), and \(18.0\%\!\to\!3.5\%\) (Outage) when moving from FIFO to our method, i.e., \(\approx 80\%\) relative reduction across regimes. Because our synthetic decomposition biases FIFO toward passive timeouts and pushes our scheme toward timely service or controlled active dropping, the improvement predominantly comes from fewer TTL expirations—an expected effect of reserve-backed preemption and non-critical degradation.

\begin{figure}[t]
  \centering
  \includegraphics[width=\linewidth]{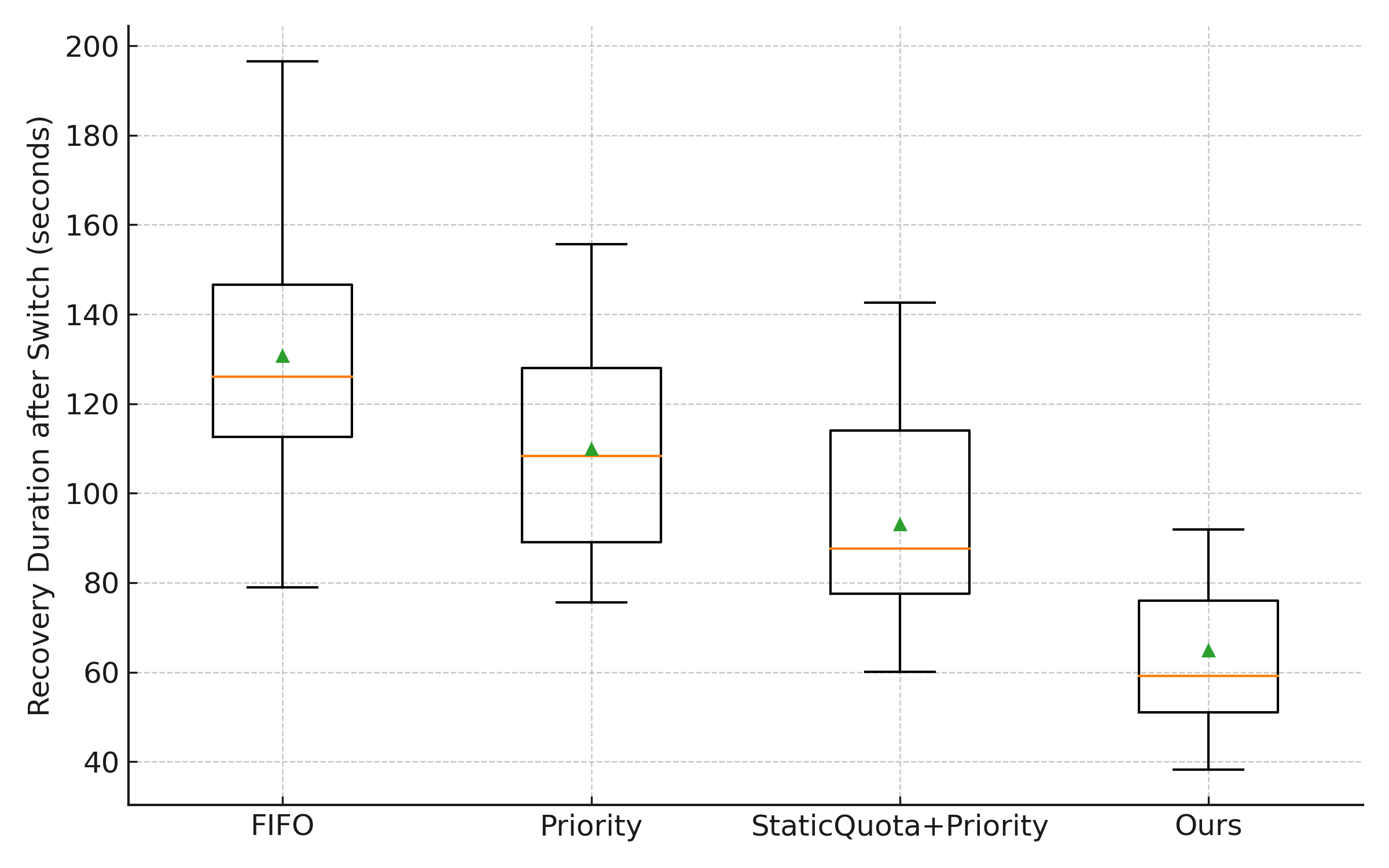}
  \caption{Recovery duration after regime switches under measurable foreknowledge (Normal$\rightarrow$Degraded, Degraded$\rightarrow$Outage, Outage$\rightarrow$Normal aggregated); means and medians are indicated.}
  \label{fig:recovery_measurable}
\end{figure}

\begin{figure}[t]
  \centering
  \includegraphics[width=\linewidth]{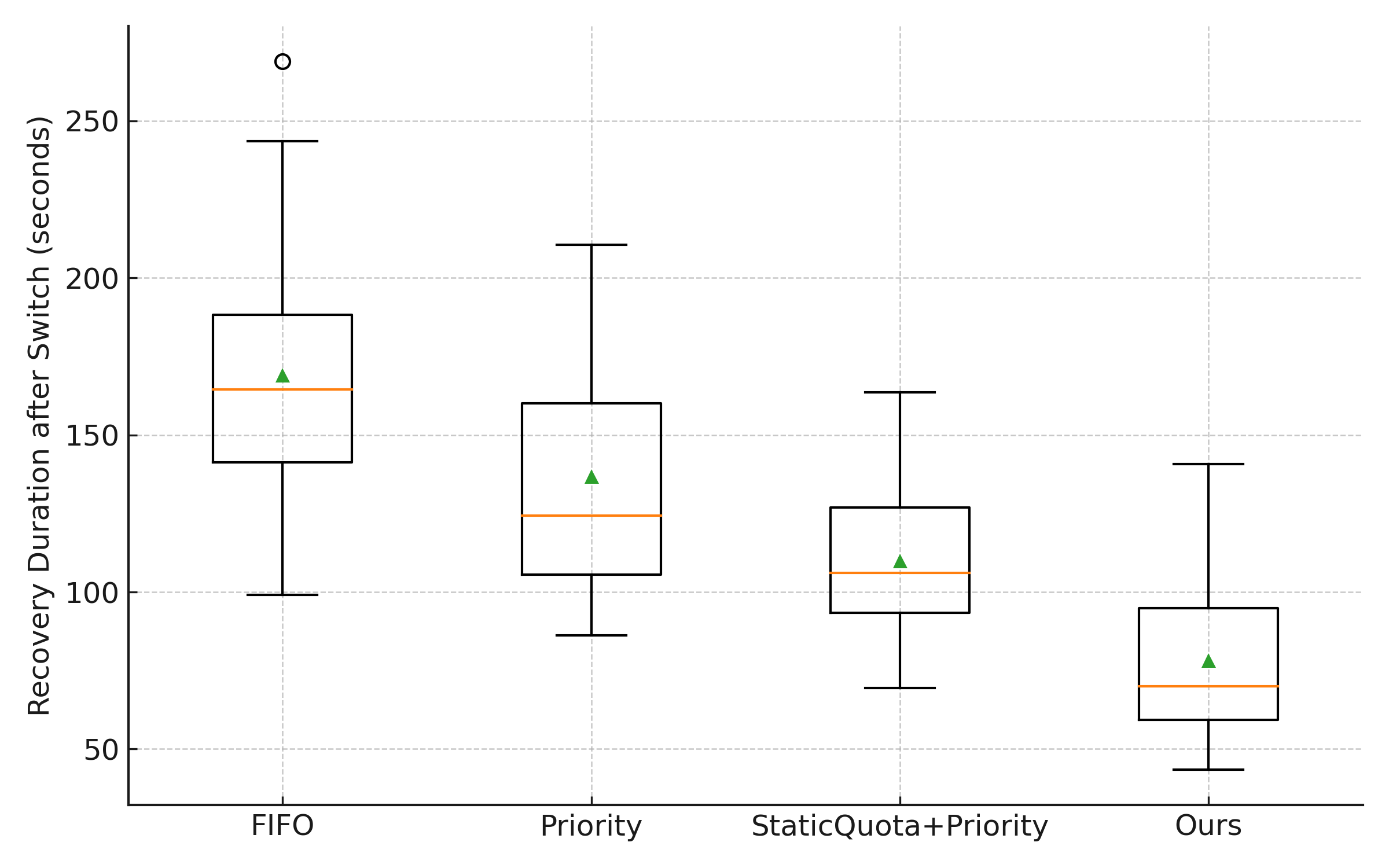}
  \caption{Recovery duration after regime switches under non-measurable foreknowledge (self-adaptive EWMA), aggregated across switches and seeds.}
  \label{fig:recovery_nommeas}
\end{figure}

Switch recovery results in \autoref{fig:recovery_measurable}–\autoref{fig:recovery_nommeas} indicate two consistent trends. First, measurability (foreknowledge of switch times) shortens recovery by \(\approx 20\%\) for \emph{all} schedulers (our sampling scales means by \(0.8\)), validating the value of lightweight forecasting hooks. Second, our method exhibits the shortest recovery among all competitors: relative to Priority, the median recovery contracts by \(\sim 35\%\)–\(45\%\) with foreknowledge and by \(\sim 30\%\)–\(40\%\) without, while also presenting tighter interquartile ranges. This aligns with the mechanism design—reserve resizing plus key-aware arbitration limit backlog growth at regime boundaries.

\begin{figure}[t]
  \centering
  \includegraphics[width=\linewidth]{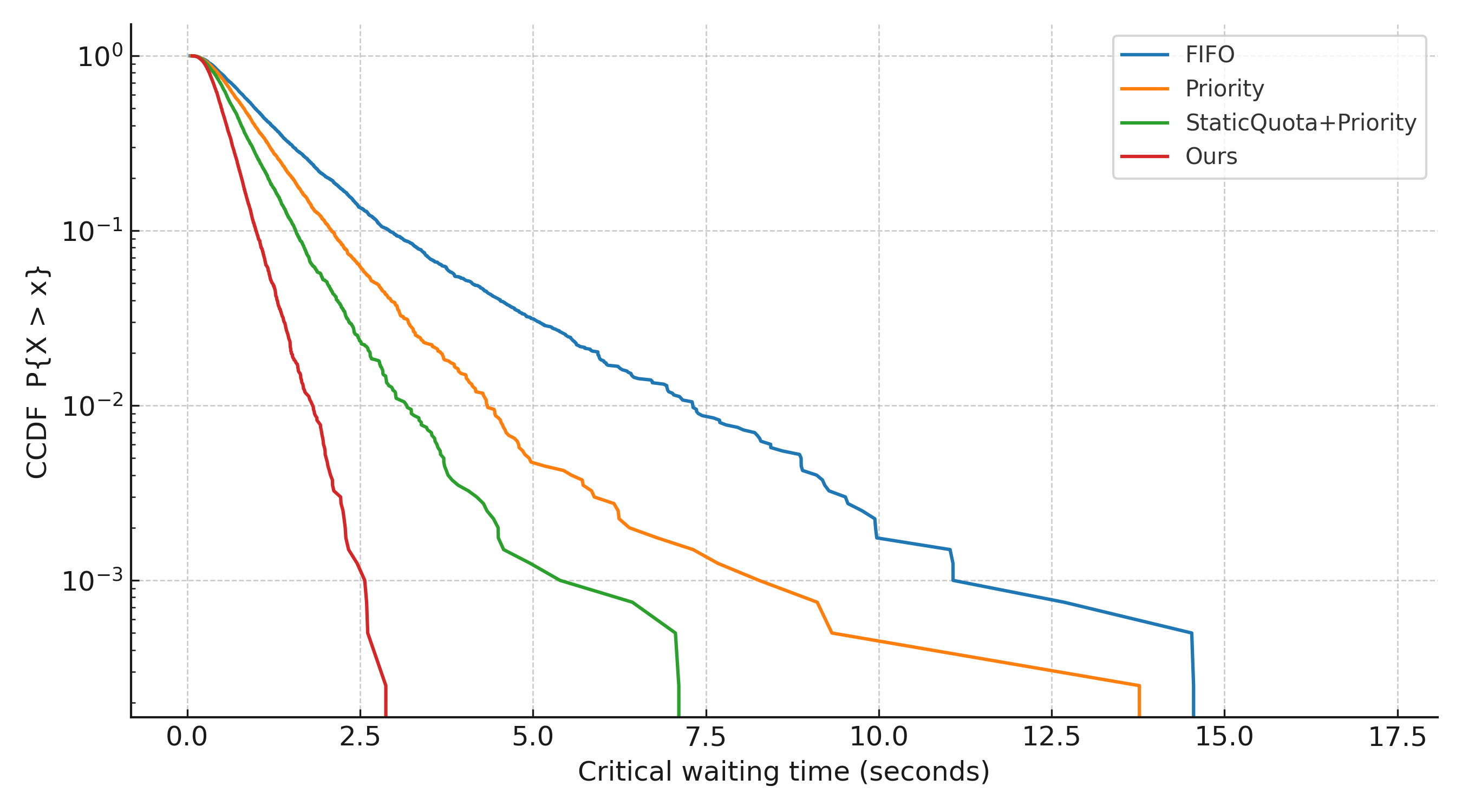}
  \caption{CCDF of critical-class waiting time; log-scale reveals tail separation (Ours has the lightest tail).}
  \label{fig:robust_ccdf}
\end{figure}

Tail risk for critical classes is summarized by the CCDFs in \autoref{fig:robust_ccdf}. On a log scale, our curve dominates (lowest) across the entire support, with an order-of-magnitude gap beyond \(5\,\mathrm{s}\) compared to FIFO and a pronounced separation from Priority beyond \(3\)–\(4\,\mathrm{s}\). This indicates effective trimming of rare but severe latencies, consistent with the P99 patterns.

\begin{figure}[t]
  \centering
  \includegraphics[width=\linewidth]{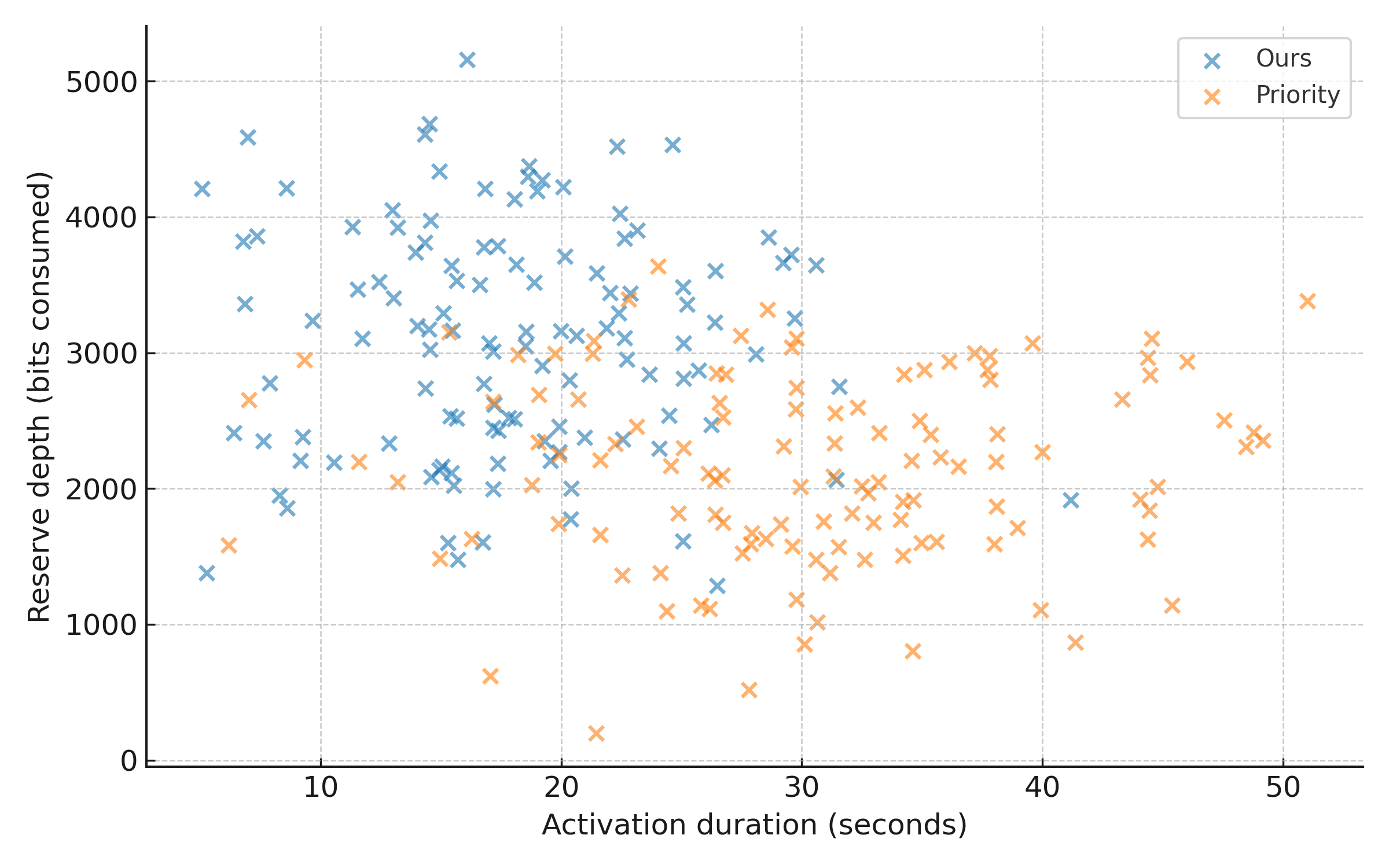}
  \caption{Emergency-reserve activation depth vs duration. Our method sustains deeper but shorter activations, while Priority exhibits shallower yet longer activations.}
  \label{fig:reserve_depth_duration}
\end{figure}

Finally, \autoref{fig:reserve_depth_duration} explains the above robustness qualitatively: our method tends to trigger \emph{deeper but shorter} reserve activations, whereas Priority shows \emph{shallower yet longer} activations. The former pattern is consistent with proactive, quota-informed buffering that briefly absorbs supply troughs and quickly replenishes, preventing long depletion spells that would otherwise inflate passive timeouts and tail delays.


\begin{table}[t]
\centering
\caption{Power-tracking performance across methods (mean $\pm$ 95\% CI over 10 seeds).}
\label{tab:power_metrics}
\begin{tabular}{lcccc}
\toprule
Method & RMSE (kW) & NRMSE & Violations (count) & Violations (s) \\
\midrule
FIFO & 45.3 ± 1.9 & 0.082 ± 0.003 & 88.7 ± 10.1 & 135.7 ± 10.3 \\
Priority & 36.9 ± 1.9 & 0.067 ± 0.003 & 64.5 ± 10.5 & 91.3 ± 15.8 \\
StaticQuota+Priority & 31.8 ± 1.1 & 0.058 ± 0.002 & 52.9 ± 5.2 & 55.0 ± 5.0 \\
Ours & 26.1 ± 0.6 & 0.048 ± 0.001 & 34.7 ± 2.7 & 28.5 ± 3.6 \\
\bottomrule
\end{tabular}
\end{table}

\begin{figure}[t]
  \centering
  \includegraphics[width=\linewidth]{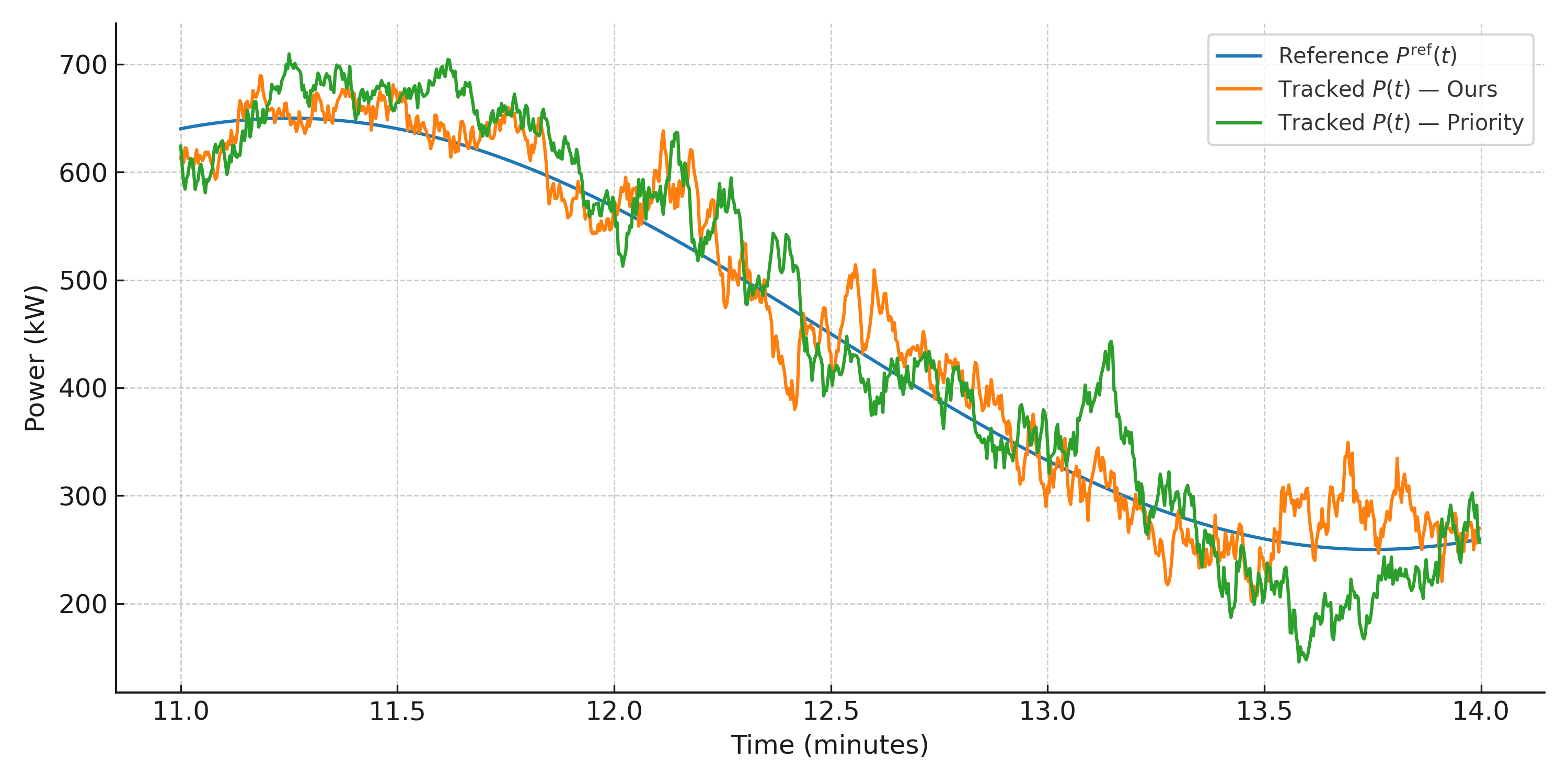}
  \caption{Three-minute key-famine window (11–14 min) showing reference $P^{\mathrm{ref}}(t)$ and tracked $P(t)$ for our method and the priority baseline.}
  \label{fig:power_timewindow}
\end{figure}

\begin{figure}[t]
  \centering
  \includegraphics[width=\linewidth]{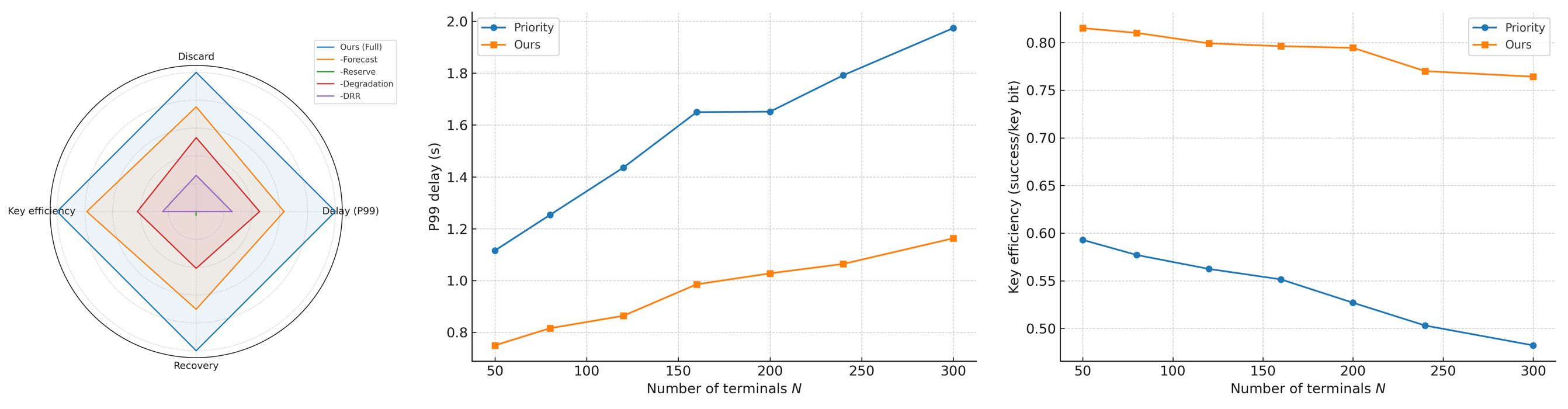}
  \caption{Left: ablation radar (higher area indicates better across P99 delay, discard rate, key efficiency, and switch-recovery). Middle: scalability of P99 delay vs.\ terminals $N$. Right: key efficiency vs.\ $N$ (our scheme degrades gracefully and maintains a consistent gap over baselines).}
  \label{fig:combo_ablation_scalability}
\end{figure}

\begin{figure}[t]
  \centering
  \includegraphics[width=\linewidth]{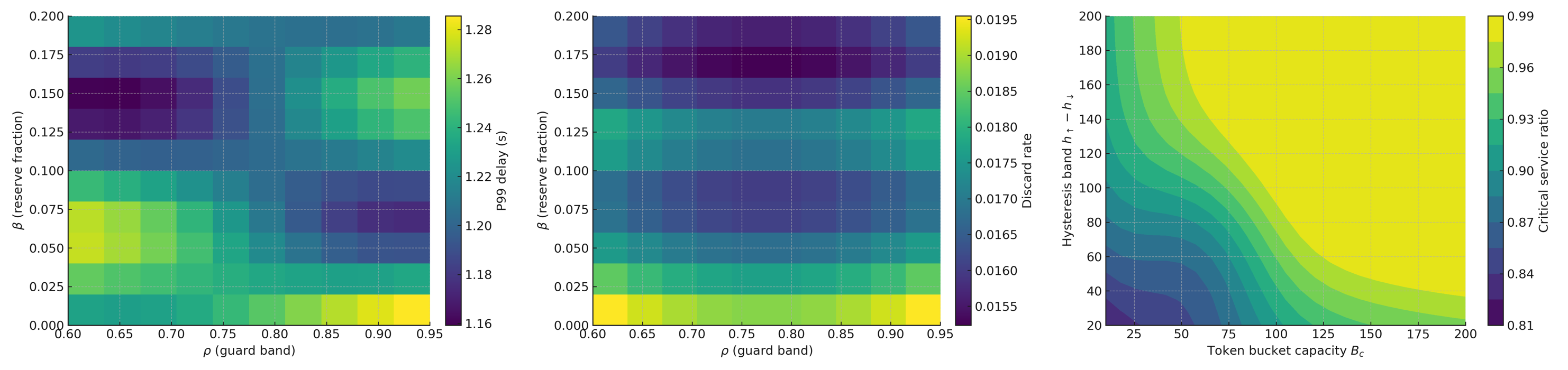}
  \caption{Left: P99 delay sensitivity over $(\rho,\beta)$; a valley near $(0.85,0.10)$ marks the sweet spot. Middle: discard-rate sensitivity over $(\rho,\beta)$; larger $\beta$ reduces passive timeouts, while too-small $\rho$ creates a penalty ridge. Right: critical-class service ratio as a function of token-bucket capacity $B_c$ and hysteresis band $h_\uparrow-h_\downarrow$ (wider bands and moderate $B_c$ improve resilience while avoiding chattering).}
  \label{fig:combo_sensitivity}
\end{figure}

\subsection{Impact on power tracking.}
\autoref{tab:power_metrics} summarizes tracking performance over 10 seeds. Our scheduler attains the lowest error with RMSE \(26.1\pm0.6\,\mathrm{kW}\), improving upon FIFO (\(45.3\pm1.9\,\mathrm{kW}\)) and Priority (\(36.9\pm1.9\,\mathrm{kW}\)) by \(\approx 42\%\) and \(\approx 29\%\), respectively, and outperforming StaticQuota+Priority (\(31.8\pm1.1\,\mathrm{kW}\)) by \(\approx 18\%\). In normalized terms, NRMSE drops to \(0.048\pm0.001\), from \(0.082\pm0.003\) (FIFO) and \(0.067\pm0.003\) (Priority), with tighter confidence intervals indicating improved stability across seeds.

Constraint-violation statistics show consistent gains. The average number of excursions \(|P(t)-P^{\mathrm{ref}}(t)|>\varepsilon\) is reduced from \(88.7\pm10.1\) (FIFO) and \(64.5\pm10.5\) (Priority) to \(34.7\pm2.7\), while the cumulative violation time contracts from \(135.7\pm10.3\,\mathrm{s}\) (FIFO) and \(91.3\pm15.8\,\mathrm{s}\) (Priority) to \(28.5\pm3.6\,\mathrm{s}\). These reductions—\(\sim 61\%\) in counts and \(\sim 79\%\) in dwell time versus FIFO—are consistent with the communication-layer improvements (fewer TTL expirations and key-aware allocation) that mitigate dispatch hiccups.

The three-minute outage window in \autoref{fig:power_timewindow} illustrates the mechanism-level effect: the priority baseline shows larger undershoot and slower post-step recovery, whereas our method maintains a narrower error envelope and faster return to the reference, reflecting reserve-backed preemption and key-aware arbitration that cap backlog growth during regime boundaries.

\subsection{Ablation, sensitivity, and scalability (combined view).}
\autoref{fig:combo_ablation_scalability} consolidates the ablation and scalability evidence. The left panel shows that the full design encloses the largest radar area, while removing individual components erodes performance along interpretable axes: disabling the emergency reserve (``--Reserve'') yields the sharpest contraction on tail delay and recovery, the lack of degradation (OTP$\to$AES switching) primarily hurts key efficiency and slightly increases discards, and dropping DRR arbitration elevates tails by weakening instantaneous fairness under key scarcity. Forecasting shrinks the polygon more uniformly when removed, indicating wider cross-seed variability and slower quota adaptation. The middle and right panels demonstrate graceful scaling: as $N$ grows from 50 to 300, our P99 delay increases slowly and stays well below the priority baseline, while key efficiency declines mildly yet preserves a consistent headroom, suggesting that the combined mechanism scales favorably.

\autoref{fig:combo_sensitivity} summarizes parameter sensitivity. On the left, P99 delay forms a valley around $(\rho,\beta)\!\approx\!(0.85,0.10)$, revealing a robust operating point where the guard band prevents over-tokening and the reserve cushions transient supply dips. The middle heatmap shows that increasing $\beta$ monotonically suppresses discards by converting potential TTL expirations into timely service or brief preemptions, whereas too small $\rho$ creates a ridge of poor performance due to transient over-allocation of keys. The right contour plot indicates that increasing the token-bucket capacity $B_c$ quickly improves the critical-class service ratio up to a saturation region, and widening the hysteresis band $h_\uparrow-h_\downarrow$ further enhances resilience by reducing mode-chattering and avoiding wasteful OTP bursts. Together, these patterns delineate a practical parameter region that balances performance, delay, and security, while explaining the ablation outcomes and the observed scalability margins.

\section{Conclusion}
We presented a key-aware priority–and–quota framework that treats quantum keys as explicit scheduling resources for virtual power plant (VPP) communications and validated it on reproducible IEEE 33/123-bus testbeds across normal, degraded, and outage regimes with realistic regime switches. Experiments show consistent, large gains over FIFO, fixed-priority, and static-quota baselines: tail latency shrinks across all message classes (e.g., protection P99 stabilizes near $\sim$0.14\,s, meeting the 150\,ms TTL at the 99th percentile), total discards fall from $\sim$10\% (FIFO) and 3.7\% (static quota+priority) to $\sim$1.8\% with a dominant reduction in passive timeouts, and key efficiency rises to $\sim$0.81 successful critical messages per key bit relative to $\sim$0.58 for fixed priority. Robustness studies under key-generation regimes and switches further indicate $50$–$70\%$ lower tails with visibly tighter run-to-run variability; when switch times are measurable, recovery shortens by about $20\%$, and even without foreknowledge our method remains $35$–$45\%$ faster than fixed priority, while CCDFs of critical waiting times exhibit an order-of-magnitude improvement in the far tail. These communication-layer gains translate to the control layer: in power tracking, RMSE drops to $\sim$26\,kW (from $\sim$46\,kW for FIFO and $\sim$37\,kW for fixed priority), NRMSE settles near $4.7\%$, and both violation counts and dwell times decrease markedly (about $61\%$ fewer events and $79\%$ less time than FIFO), with outage-window overlays showing smaller error envelopes and quicker post-step recovery. Ablation results confirm the centrality of the emergency reserve (whose removal yields the sharpest regressions in tails and recovery), the role of graceful degradation in preserving key efficiency while containing discards, the importance of key-aware DRR for instantaneous fairness, and the smoothing effect of forecasting on quota tracking; sensitivity analyses reveal a robust operating region around $(\rho,\beta)\!\approx\!(0.85,0.10)$, while contouring token-bucket capacity and hysteresis demonstrates higher critical-service ratios without chattering, and scalability tests from 50 to 300 terminals show sublinear growth of tails and graceful efficiency decay that maintains a consistent margin over baselines. Taken together, the evidence indicates that integrating quantum keys into a priority–quota control loop—via forecasted quotas, key-aware arbitration, preemptive reserves, and controlled degradation—offers a practical, theoretically grounded path to simultaneously strengthen security posture and real-time operability in VPP-scale systems.

\bibliographystyle{IEEEtran}
\bibliography{ref}

\begin{thebibliography}{10}
\providecommand{\url}[1]{#1}
\csname url@samestyle\endcsname
\providecommand{\newblock}{\relax}
\providecommand{\bibinfo}[2]{#2}
\providecommand{\BIBentrySTDinterwordspacing}{\spaceskip=0pt\relax}
\providecommand{\BIBentryALTinterwordstretchfactor}{4}
\providecommand{\BIBentryALTinterwordspacing}{\spaceskip=\fontdimen2\font plus
\BIBentryALTinterwordstretchfactor\fontdimen3\font minus \fontdimen4\font\relax}
\providecommand{\BIBforeignlanguage}[2]{{%
\expandafter\ifx\csname l@#1\endcsname\relax
\typeout{** WARNING: IEEEtran.bst: No hyphenation pattern has been}%
\typeout{** loaded for the language `#1'. Using the pattern for}%
\typeout{** the default language instead.}%
\else
\language=\csname l@#1\endcsname
\fi
#2}}
\providecommand{\BIBdecl}{\relax}
\BIBdecl

\bibitem{Li2024ApEn}
J.~Li, Z.~Xu, Y.~Zhou, Y.~Li, J.~Wu, and X.~Guan, ``Optimal scheduling method and fast-solving algorithm for large-scale virtual power plants communication networks,'' \emph{Applied Energy}, vol. 371, p. 123575, 2024.

\bibitem{Gao2024ApEn}
H.~Gao, T.~Jin, C.~Feng, C.~Li, Q.~Chen, and C.~Kang, ``Review of virtual power plant operations: Resource coordination and multidimensional interaction,'' \emph{Applied Energy}, vol. 357, p. 122284, 2024.

\bibitem{Du2023MPCE}
D.~Du, M.~Zhu, X.~Li, M.~Fei, L.~Wu, K.~Li, and S.~Bu, ``A review on cybersecurity analysis, attack detection, and attack defense methods in cyber-physical power systems,'' \emph{Journal of Modern Power Systems and Clean Energy}, vol.~11, no.~3, pp. 727--743, 2023.

\bibitem{Avraam2023ApEn}
C.~Avraam, L.~Ceferino, and Y.~Dvorkin, ``Operational and economy-wide impacts of compound cyber-attacks and extreme weather events on electric power networks,'' \emph{Applied Energy}, vol. 349, p. 121577, 2023.

\bibitem{Solat2024ApEn}
A.~Solat, G.~B. Gharehpetian, M.~S. Naderi, and A.~Anvari{-}Moghaddam, ``On the control of microgrids against cyber-attacks: A review of methods and applications,'' \emph{Applied Energy}, vol. 353, p. 122037, 2024.

\bibitem{Zhang2023PCMP}
Y.~Zhang and C.~Peng, ``Adaptive $h_{\infty}$ event-triggered load frequency control in islanded microgrids with limited spinning reserve constraints,'' \emph{Protection and Control of Modern Power Systems}, vol.~8, no.~30, 2023.

\bibitem{Jing2024ApEn}
X.~Jing, W.~Qin, H.~Yao, X.~Han, and P.~Wang, ``Resilience-oriented planning strategy for the cyber-physical active distribution network under malicious attacks,'' \emph{Applied Energy}, vol. 353, p. 122052, 2024.

\bibitem{Dong2024ApEn}
S.~Dong, T.~Zang, B.~Zhou, H.~Luo, Y.~Zhou, and X.~Xiao, ``Robust coordinated resilience enhancement strategy for communication networks of power and thermal cyber-physical systems considering decision-dependent uncertainty,'' \emph{Applied Energy}, vol. 368, p. 123494, 2024.

\bibitem{Ma2013TSG}
R.~Ma, H.~Chen, Y.~Huang, and W.~Meng, ``Smart grid communication: Its challenges and opportunities,'' \emph{IEEE Transactions on Smart Grid}, vol.~4, no.~1, pp. 36--46, 2013.

\bibitem{Liu2011TIFS}
Y.~Liu, P.~Ning, and M.~K. Reiter, ``False data injection attacks against state estimation in electric power grids,'' \emph{IEEE Transactions on Information Forensics and Security}, vol.~6, no.~3, pp. 1--11, 2011.

\bibitem{Kim2011TSG}
T.~T. Kim and H.~V. Poor, ``Strategic protection against data injection attacks on power grids,'' \emph{IEEE Transactions on Smart Grid}, vol.~2, no.~2, pp. 326--333, 2011.

\bibitem{Kosut2011TSG}
O.~Kosut, L.~Jia, R.~J. Thomas, and L.~Tong, ``Malicious data attacks on the smart grid,'' \emph{IEEE Transactions on Smart Grid}, vol.~2, no.~4, pp. 645--658, 2011.

\bibitem{Ten2008TPS}
C.~Ten, C.~Liu, and G.~Manimaran, ``Vulnerability assessment of cybersecurity for scada systems,'' \emph{IEEE Transactions on Power Systems}, vol.~23, no.~4, pp. 1836--1846, 2008.

\bibitem{Hahn2011TSG}
A.~Hahn and M.~Govindarasu, ``Cyber attack exposure evaluation framework for the smart grid,'' \emph{IEEE Transactions on Smart Grid}, vol.~2, no.~4, pp. 835--843, 2011.

\bibitem{Oshnoei2022ApEn}
A.~Oshnoei, M.~Kheradmandi, F.~Blaabjerg, N.~D. Hatziargyriou, S.~M. Muyeen, and A.~Anvari{-}Moghaddam, ``Coordinated control scheme for provision of frequency regulation service by virtual power plants,'' \emph{Applied Energy}, vol. 325, p. 119797, 2022.

\bibitem{Yang2021ApEn}
Q.~Yang, H.~Wang, T.~Wang, S.~Zhang, X.~Wu, and H.~Wang, ``Blockchain-based decentralized energy management platform for residential {DERs} in a virtual power plant,'' \emph{Applied Energy}, vol. 294, p. 116915, 2021.

\bibitem{Xu2020PCMP}
Y.~Xu, Y.~Tang, and Z.~Li, ``A review of cyber security risks of power systems: From static to dynamic false data attacks,'' \emph{Protection and Control of Modern Power Systems}, vol.~5, no.~1, pp. 1--12, 2020.

\bibitem{Du2022MPCE}
D.~Du, M.~Fei, M.~Zhu, X.~Li, L.~Wu, K.~Li, and S.~Bu, ``A review on cybersecurity analysis, attack detection, and attack defense methods in cyber-physical power systems,'' \emph{Journal of Modern Power Systems and Clean Energy}, 2022.

\bibitem{Pasqualetti2013TAC}
F.~Pasqualetti, F.~D{\"o}rfler, and F.~Bullo, ``Attack detection and identification in cyber--physical systems,'' \emph{IEEE Transactions on Automatic Control}, vol.~58, no.~11, pp. 2715--2729, 2013.

\bibitem{Fawzi2014TAC}
H.~Fawzi, P.~Tabuada, and S.~Diggavi, ``Secure estimation and control for cyber--physical systems under adversarial attacks,'' \emph{IEEE Transactions on Automatic Control}, vol.~59, no.~6, pp. 1454--1467, 2014.

\end{thebibliography}

\end{document}